\documentclass[preprintnumbers,amsmath,amssymb,prd]{revtex4}
\usepackage{graphicx}
\usepackage{multibox}

\def\bbox{{\,\lower0.9pt\vbox{\hrule \hbox{\vrule height 0.2 cm\hskip
0.2 cm \vrule height 0.2 cm}\hrule}\,}}
\newcommand{\dsl}{\pa \kern-0.5em /}
\renewcommand{\t}{\theta}

\newcommand{\nn}{\nonumber \\}

\def\ben{\begin{equation}}
\def\een{\end{equation}}
\def\bena{\begin{eqnarray}}
\def\eena{\end{eqnarray}}

\newcommand{\be}{\begin{equation}}
\newcommand{\ee}{\end{equation}}
\newcommand{\bea}{\begin{eqnarray}}
\newcommand{\eea}{\end{eqnarray}}
\newcommand{\beq}{\begin{eqnarray}}
\newcommand{\eeq}{\end{eqnarray}}
\def\e{\epsilon}
\def\6{\partial}

\def\a{\alpha}
\def\g{\gamma}
\def\d{\delta}
\def\b{\beta}

\def\m{\mu}
\def\n{\nu}
\def\r{\rho}
\def\l{\lambda}
\def\s{\sigma}

\def\pa{\partial}
\def\B{\begin{equation}}
\def\E{\end{equation}}

\def\today{\ifcase\month\or  January\or February\or March\or April\or
May\or June\or  July\or August\or September\or October\or
November\or December\fi \space\number\day, \number\year}
\input amssym.def
\input amssym.tex

\begin{document}
\preprint{CECS-PHY-06/03}
 \preprint{NSF-KITP-06-03}
\preprint{ULB-TH/06-02}
\title{Monopoles for Gravitation and for Higher Spin Fields}
\author{Claudio Bunster}
\affiliation{Centro de Estudios Cient\'{\i}ficos (CECS), Valdivia,
Chile.}
\author{Sandrine Cnockaert}
\affiliation{Physique Th\'eorique et Math\'ematique, Universit\'e
Libre de Bruxelles \& International Solvay Institutes, ULB Campus
Plaine C.P. 231, B--1050 Bruxelles, Belgium.}
\author{Marc Henneaux}
\affiliation{Physique Th\'eorique et Math\'ematique, Universit\'e
Libre de Bruxelles \& International Solvay Institutes, ULB Campus
Plaine C.P. 231, B--1050 Bruxelles, Belgium, and \\Centro de
Estudios Cient\'{\i}ficos (CECS), Valdivia, Chile.}
\author{Rub\'en Portugues}
\affiliation{&&Centro de Estudios Cient\'{\i}ficos (CECS),
Valdivia, Chile.}

\begin{abstract}

We consider massless higher spin gauge theories with both electric
and magnetic sources, with a special emphasis on the spin two case.
We write the equations of motion at the linear level (with conserved
external sources) and introduce Dirac strings so as to derive the
equations from a variational principle.  We then derive a
quantization condition that generalizes the familiar Dirac
quantization condition, and which involves the conserved charges
associated with the asymptotic symmetries for higher spins.  Next we
discuss briefly how the result extends to the non linear theory.
This is done in the context of gravitation, where the Taub-NUT
solution provides the exact solution of the field equations with
both types of sources. We rederive, in analogy with
electromagnetism, the quantization condition from the quantization
of the angular momentum. We also observe that the Taub-NUT metric is
asymptotically flat at spatial infinity in the sense of Regge and
Teitelboim (including their parity conditions). It follows, in
particular, that one can consistently consider in the variational
principle configurations with different electric and magnetic
masses.

\end{abstract}

\maketitle

\section{Introduction}

The idea of electric-magnetic duality has resiliently resisted
developments and appears to be here to stay.  Originally explored by
Dirac for electromagnetism \cite{Dirac:1931kp,Dirac:1948um}, it was
later analyzed in the context of non-abelian gauge theories in
\cite{'tHooft:1974qc,Polyakov:1974ek,Montonen:1977sn}. It was more
recently generalized to extended objects and $p$-form gauge fields
in \cite{CT,RN}.  A fascinating implication of electric-magnetic
duality is the charge quantization condition. This condition is
antisymmetric for p-dyons of even spatial dimension p, and symmetric
for odd p \cite{Schwinger:1968rq,Zwanziger:1968rs,Deser:1997mz}.

{}Fields with spin higher than that of the graviton were discussed
some time ago by Fronsdal \cite{Fronsdal:1978rb} who gave the action
integral for the free theories (see also \cite{BdWFC}).  Interest in
these higher spin fields has been renewed since they appear in the
spectrum of string theory. In particular, duality for free higher
spins has recently
been the focus of many papers \cite{All,Francia:2002pt,%
Bekaert:2003az,Boulanger,Henneaux:2004jw,Julia:2005ze}. Conserved,
external, electric-type sources can easily be coupled to the
higher spin fields but the problem of constructing consistent
self-interactions is still  incompletely understood beyond spin 2 (for instance, the action principle is not known even though non linear equations have been constructed in the remarkable work \cite{Vasiliev:1990en}). It has been argued
that self-interactions can be brought in but only among an
infinite tower of fields \cite{Vasiliev:1990en}.

Since duality can be defined for higher spins and since electric
sources can be included, one might wonder whether magnetic sources
can be considered as well. Our paper solves positively this question
for all spins at the linearized level and provides additional
insight in the full non linear theory for spin 2.

We show that conserved external sources of both types can be coupled
to any given higher (integer) spin field within the context of the
linear theory.  The presence of magnetic sources requires the
introduction of Dirac strings, as in the spin 1 case. To preserve
manifest covariance, the location of the string must be left
arbitrary and is, in fact, unobservable classically.  The
requirement that the Dirac string is unobservable
quantum-mechanically forces a quantization condition of the form \be
\frac{1  }{2 \pi \hbar} \, Q_{\g_1 \cdots \g_{s-1}}(v)P^{\g_1 \cdots
\g_{s-1}}(u) \in \Bbb{Z} \,.\ee  Here, the symmetric tensor $P^{\g_1
\cdots \g_{s-1}}(u)$ is the conserved electric charge associated
with the asymptotic symmetries of the spin s field, while $Q_{\g_1
\cdots \g_{s-1}}(v)$ is the corresponding ``topological" magnetic
charge. For s = 1, the asymptotic symmetries are internal symmetries
and, actually, just constant phase transformations. The conserved
charge $P$ is the electric charge $q$ while $Q$ is the magnetic
charge $g$, yielding the familiar Dirac quantization condition for
the product of electric and magnetic charges.  {}For s = 2 the
conserved charges have a spacetime index and the quantization
condition reads (after rescaling the conserved quantities so they
have dimensions of mass) \be \frac{4 G P_\g Q^\g}{\hbar} \in \Bbb{Z}
\, . \label{QC1}\ee The quantity $P_\g$ is the ``electric"
4-momentum associated with constant linearized diffeomorphisms
(translations) while $Q_\g$ is the corresponding magnetic
4-momentum.  {}For a point particle source, $P_\g = M u_\g$ where
$M$ is the ``electric" mass and $u_\g$ the 4-velocity of the
electric source. Similarly, $Q_\g = N v_\g$ where $N$ is the
``magnetic" mass and $v_\g$ the 4-velocity of the magnetic source.

All this is just a generalization of the familiar spin 1 case,
although the explicit introduction of the Dirac string is more
intricate for higher spins because the gauge invariance is then more
delicate to control.  Indeed, there is no gauge invariant object
that involves first derivatives of the fields only (s $>$ 1). Hence,
the Lagrangian is not strictly gauge invariant, contrary to what
happens for electromagnetism, but is gauge invariant only up to a
total derivative.

A serious limitation of the linear theory for s $>$ 1 is that the
sources must follow straight lines.  This follows from the strict
conservation laws implied by the field equations, which are much
more stringent for s $>$ 1 than they are for s = 1. Thus the sources
must be treated as externally given and cannot be freely varied in
the variational principle. One cannot study the back reaction of the
spin s field on the sources without introducing self-interactions.
This problem occurs already for the spin 2 case and has nothing to
do with the introduction of magnetic sources.

We have not investigated the backreaction problem for general spins
s $>$ 2 since the non linear theory is still a subject of investigation
even in the absence
of sources. We have focused instead on the spin 2 case, for which
the non linear theory is given by the Einstein theory of gravity.
The remarkable Taub-NUT solution \cite{Newman:1963yy}, which
represents the vacuum exterior field of a gravitational dyon,
indicates that Einstein's theory can support both electric and
magnetic masses. The remaining part of our paper is devoted to a
further discussion of the Taub-NUT solution, including an analysis
of its asymptotic properties.  We provide new insight on the
quantization condition (\ref{QC1}) through the analysis of the
angular momentum stored in the gravitational field of a pair of
electric and magnetic masses placed at different locations.

Our paper is organized as follows. In section II, we consider in
detail the spin 2 case with point particle electric and magnetic
sources. We introduce Dirac strings and derive the quantization
condition. We then extend the formalism to higher spins (section
\ref{higherspins0}), again with point particle sources. In section
\ref{sect4} we consider the Taub-NUT solution and discuss the
quantization condition in the non linear context. After a brief
concluding section, the article ends with three appendices where
technical aspects for spins greater than 2 that were not included in
section \ref{higherspins0} are dealt with (Appendices
\ref{appendixA}, \ref{appendixGI} and \ref{QCHS}).  Finally some
aspects of the angular momentum of an electric test particle in the
field of a monopole, which turn out to be relevant to the discussion
of the Taub-NUT metric are covered in Appendix \ref{appendixB}.

\section{Linearized gravity with electric and magnetic masses}
\setcounter{equation}{0}

\subsection{Electric and magnetic sources}

The equations of motion for linearized gravity coupled to both
electric and magnetic sources are naturally written in terms of
the linearized Riemann tensor $R_{\a \b \l \m}$, hereafter just
called ``Riemann tensor" for simplicity. This is the physical,
gauge-invariant, object, analogous to the field strength $F_{\m
\n}$ of electromagnetism.  How to introduce the ``potential",
i.e., the symmetric spin 2 field $h_{\m \n} = h_{\n \m}$ will be
discussed below.  The dual to the Riemann tensor is defined as $$
S_{\a \b \l \m} = - \frac{1}{2} \epsilon_{\a \b \g \d} R^{\g
\d}_{\;\;\;\; \l \m}.$$

We denote the ``electric" energy-momentum tensor by $T^{\m \n}$ and
the ``magnetic" energy-momentum tensor by $\Theta^{\m \n}$. These
are both symmetric and conserved, $T^{\m \n} = T^{\n \m}$,
$\Theta^{\m \n} = \Theta^{\n \m}$, $ T^{\m \n}_{\; \; \; \; , \, \n}
= 0$, $ \Theta^{\m \n} _{\; \; \; \; , \, \n} = 0 $. It is also
useful to define $ \bar{T}^{\m \n} = T^{\m \n} - \frac{1}{2} \,
\eta^{\m \n}\, T $, $ \bar{\Theta}^{\m \n} = \Theta^{\m \n} -
\frac{1}{2} \, \eta^{\m \n}\, \Theta $ where $T$ and $\Theta$ are
the traces. We assume that $T^{\m \n}$ and $\Theta^{\m \n}$ have the
units of an energy density.  We set $c=1$ but keep $G$.

The form of the equations in the presence of both types of sources
is fixed by: (i) requiring duality invariance with respect to the
$SO(2)$-rotations of the curvatures and the sources
\cite{Henneaux:2004jw},  \beq R'_{\a \b \l \m}  &=& \cos \a \, R_{\a
\b \l \m} + \sin \a \, S_{\a \b \l \m} , \;\; \; \; S'_{\a \b \l \m}
= - \sin \a \, R_{\a \b \l \m} + \cos \a \, S_{\a \b \l \m},
\nonumber\\ T'_{\a \b} &=& \cos \a \, T_{\a \b} + \sin \a \,
\Theta_{\a \b} , \; \; \; \;\; \; \; \; \; \; \; \; \; \; \; \;
\Theta'_{\a \b} = - \sin \a \, T_{\a \b} + \cos \a \, \Theta_{\a
\b}, \nonumber \eeq and, (ii) using the known form of the equations
in the presence of electric masses only.  One finds explicitly the
following:
\begin{itemize}
\item The Riemann tensor is antisymmetric in the first two indices
and the last two indices, but in general is not symmetric for the
exchange of the pairs, i.e., $R_{\a \b \l \m} = - R_{\b \a \l
\m}$, $R_{\a \b \l \m} = - R_{\a \b \m \l}$ with $R_{\a \b \l \m}
\not= R_{\l \m \a \b }$ (in the presence of magnetic sources).
\item In the presence of
magnetic sources the cyclic identity is \footnote{In terms of the
Riemann tensor, this ``identity" is a non trivial equation and not
an identity. It becomes an identity only after the Riemann tensor is
expressed in terms of the spin 2 field $h_{\m \n}$ introduced below.
We shall nevertheless loosely refer to this equation as the
(generalized) cyclic identity. A similar remark holds for the
Bianchi identity below.} \be R_{\a \b \l \m} + R_{\b \l \a \m} +
R_{\l \a \b \m} = 8 \pi G\, \e_{\a \b \l \n} \,
\bar{\Theta}^{\n}_{\; \; \m} \label{cyclic}\,.\ee This enables one
to relate $R_{\a \b \l \m}$ to $R_{\l \m \a \b }$ through \be R_{\a
\b \g \d} - R_{\g \d \a \b} = 4 \pi G \left( \e_{\a \b \g \l}
\bar{\Theta}^\l_{\; \; \d} - \e_{\a \b \d \l} \bar{\Theta}^\l_{\; \;
\g} + \e_{\b \g \d \l} \bar{\Theta}^\l_{\; \; \a} - \e_{\a \g \d \l}
\bar{\Theta}^\l_{\; \; \b} \right). \label{symm}\ee It follows that
the Ricci tensor is symmetric, $ R_{\l \m} = R_{\m \l} $.  The
Einstein tensor $G_{\l \m} = R_{\l \m} - (1/2) \eta_{\l \m} R $  is
then also symmetric.
\item The Bianchi identity is \be
\partial_\e R_{\a \b \g \d} +
\partial_\a R_{\b \e \g \d} + \partial_\b R_{\e \a \g \d} = 8 \pi
G \, \e_{\e \a \b \r} ( \partial_\g \bar{\Theta}^\r_{\; \; \d} -
\partial_\d \bar{\Theta}^\r_{\; \; \g} )\,. \label{BI}\ee  Although
there is now a right-hand side in the Bianchi identity, the
contracted Bianchi identities are easily verified to be unaffected
and still read \be G^{\l \m}_{\; \; \; \; , \, \m} = 0 \,.\ee \item
The Einstein equations are \be G^{\l \m} = 8 \pi G \, T^{\l \m}\,,
\label{Einstein}\ee or equivalently, $R^{\l \m} = 8 \pi G \,
\bar{T}^{\l \m}$, and force exact conservation of the sources
because of the contracted Bianchi identity, as in the absence of
magnetic mass.
\end{itemize}

The equations are completely symmetric under duality.  Indeed, one
easily checks that one gets the same equations for the dual
curvature $S_{\a \b \l \m}$ with the roles of the electric and
magnetic energy-momentum tensors exchanged.  In the course of the
verification of this property, the equation $$
\partial^\m R_{\m \r \g \d} = 8 \pi G \, \left(
\partial_\g \bar{T}_{\r \d} -
\partial_\d \bar{T}_{\r \g} \right),$$ which
follows from (\ref{symm}), (\ref{BI}) and the conservation of
$\Theta^{\m \n}$ are useful. {}Furthermore, in the absence of
magnetic sources, one recovers the equations of the standard
linearized Einstein theory since the cyclic and Bianchi identities
have no source term in their right hand sides.

The formalism can be extended to include a cosmological constant
$\Lambda$.  The relevant curvature is then the MacDowell-Mansouri
curvature \cite{McD} linearized around (anti) de Sitter space
\cite{Julia:2005ze}. In terms of this tensor, the equations
(\ref{cyclic}), (\ref{BI}) and (\ref{Einstein}) take the same
form, with ordinary derivatives replaced by covariant derivatives
with respect to the (anti) de Sitter background.

\subsection{Decomposition of the Riemann tensor - Spin 2 field}

We exhibit a variational principle from which the equations of
motion follow. To that end, we first need to indicate how to
introduce the spin 2 field $h_{\mu\nu}$.

Because there are right-hand sides in the cyclic and Bianchi
identities, the Riemann tensor is not directly derived from a
potential $h_{\m \n}$. To introduce $h_{\m \n}$, we split $R_{\l \m
\a \b}$ into a part that obeys the cyclic and Bianchi identities and
a part that is fixed by the magnetic energy-momentum tensor.  Let
$\Phi^{\a \b}_{\; \; \; \; \g}$ be such that \be \partial_\a \,
\Phi^{\a \b}_{\; \; \; \; \g} = 16 \pi G \, \Theta^\b_{\; \; \g} ,
\; \; \; \Phi^{\a \b}_{\; \; \; \; \g} = - \, \Phi^{\b \a}_{\; \; \;
\; \g} \label{phi}\,.\ee We shall construct $\Phi^{\a \b}_{\; \; \;
\; \g}$ in terms of $\Theta^\b_{\; \; \g}$ and Dirac strings below.
We set \be R_{\l \m \a \b} = r_{\l \m \a \b} + \frac{1}{4}\, \e_{\l
\m \r \s} \left(\partial_\a \bar{\Phi}^{\r \s}_{\; \; \; \; \b} -
\partial_\b \bar{\Phi}^{\r \s}_{\; \; \; \; \a} \right)\,, \label{split0}
\ee with \be \bar{\Phi}^{\r \s}_{\; \; \; \; \a} = \Phi^{\r \s}_{\;
\; \; \; \a} + \frac{1}{2} \left( \d^\r _{\; \a} \Phi^\s - \d^\s
_{\; \a} \Phi^\r \right) , \; \; \; \Phi^\r \equiv \Phi^{\r \s}_{\;
\; \; \; \s} \nonumber \,.\ee Using $ \partial_\a \, \bar{\Phi}^{\a
\b}_{\; \; \; \; \g} = 16 \pi G \, \bar{\Theta}^\b_{\; \; \g} -
\partial_\g \bar{\Phi}^\b $, $ \bar{\Phi}^\b = -
\frac{1}{2} \Phi^\b $, one easily verifies that the cyclic and
Bianchi identities take the standard form when written in terms of
$r_{\a \b \l \m}$, namely, \be r_{\a \b \l \m} + r_{\b \l \a \m} +
r_{\l \a \b \m} = 0 , \; \; \; \; \;
\partial_\e r_{\a \b \g \d} +
\partial_\a r_{\b \e \g \d} + \partial_\b r_{\e \a \g \d} = 0
.\nonumber \ee   Hence, there exists a symmetric tensor $h_{\m
\n}$ such that $ r_{\a \b \l \m} = - 2
\partial_{[\b} h_{\a][\l , \, \m ]} $.

If one sets $ y^{\l \m}_{\; \; \; \; \g} = \e^{\l \m \r \s}
\partial _\r h_{\s \g} = - y^{\m \l}_{\; \; \; \; \g}, $ one
may rewrite the curvature as \be R_{\l \m \a \b} =  \frac{1}{4}\,
\e_{\l \m \r \s} \left(\partial_\a \bar{Y}^{\r \s}_{\; \; \; \; \b}
- \partial_\b \bar{Y}^{\r \s}_{\; \; \; \; \a} \right) \,,
\label{curvature}\ee with \be \label{ybar} Y^{\r \s}_{\; \; \; \;
\b} = y^{\r \s}_{\; \; \; \; \b} + \Phi^{\r \s}_{\; \; \; \; \b} = -
Y^{\s \r}_{\; \; \; \; \b}, \; \; \; \;  \bar{Y}^{\r \s}_{\; \; \;
\; \a} = Y^{\r \s}_{\; \; \; \; \a} + \frac{1}{2} \left( \d^\r _{\;
\a} Y^\s - \d^\s _{\; \a} Y^\r \right) , \; \; \; Y^\r \equiv Y^{\r
\s}_{\; \; \; \; \s} \,,\ee (note that $\bar{y}^{\r \s}_{\; \; \; \;
\a} = y^{\r \s}_{\; \; \; \; \a}$ and that $\partial_\r y^{\r
\s}_{\; \; \; \; \a} = 0$).

\subsection{Dirac string}
We consider point particle sources.  The particles must be forced to
follow straight lines because of the conservation equations $T^{\m
\n}_{\; \; \; , \n} = 0$ and $\Theta^{\m \n}_{\; \; \; , \n} = 0$.
If $u^\m$ is the 4-velocity of the electric source and $v^\m$ the
4-velocity of the magnetic source, one has \be T^\m_{\; \; \n} = M
u_\n \int d\l \delta^{(4)}(x-z(\l)) \dot{z}^\m , \; \; \; \; \;
\Theta^\m_{\; \; \n} = N v_\n \int d\l \delta^{(4)}(x-\bar{z}(\l))
\dot{\bar{z}}^\m \,,\ee where $z^\m(\l)$ and $\bar{z}^\m(\l))$ are
the worldlines of the electric and magnetic sources respectively,
e.g. $u^\m = dz^\m /ds$. Performing the integral, one finds
$$ T^{\m\n} = \frac{u^\m u^\n}{u^0} \delta^{(3)}(\vec{x} -
\vec{z}(x^0)), \; \; \; \; \; \Theta^{\m\n} = \frac{v^\m v^\n}{v^0}
\delta^{(3)}(\vec{x} - \vec{\bar{z}}(x^0))\,.$$

The tensor $\Phi^{\a \b}_{\; \; \; \; \g}$ introduced in (\ref{phi})
can be constructed \`a la Dirac \cite{Dirac:1948um}, by attaching a
Dirac string $y^\m(\l, \s)$ to the magnetic source, $y^\m(\l,
0)=\bar{z}^\m(\l)$. One has explicitly \be \Phi^{\a \b}_{\; \; \; \;
\g} = 16 \pi G \, N v_\g \int d \l d \s (y'^\a \dot{y}^\b -
\dot{y}^\a y'^\b) \delta^{(4)}(x-y(\l, \s)) \,,\label{exprforphi}\ee
where  $$ \dot{y}^\a = \frac{\partial y^\a}{\partial \l}, \; \; \;
\; y'^\a = \frac{\partial y^\a}{\partial \s}\, .
$$ One verifies exactly as for electromagnetism that the
divergence of $\Phi^{\a \b}_{\; \; \; \; \g}$ is equal to the
magnetic energy-momentum tensor (up to the factor $16 \pi G$). What
plays the role of the magnetic charge $g$ in electromagnetism is now
the conserved product $N v_\m$ of the magnetic mass of the source by
its $4$-velocity.  This is the magnetic $4$-momentum.

\subsection{Variational principle}

\subsubsection{Action}

When the curvature is expressed in terms of $h_{\m \n}$ as in
(\ref{curvature}), the expressions (\ref{cyclic}) and (\ref{BI}) are
identically fulfilled and the relations (\ref{Einstein}) become
equations of motion for $h_{\m \n}$.  These equations can be derived
from a variational principle which we now describe.

The action that yields (\ref{Einstein}) is \be S[h_{\m
\n}(x),y^\m(\l, \s)] = \frac{1}{16 \pi G} \int \frac{1}{4}\left(
\bar{Y}_{\a \b \g} \bar{Y}^{\a \g \b} - \bar{Y}_\a
\bar{Y}^\a \right) d^4 x + \frac{1}{2} \int h_{\m \n} T^{\m \n} d^4
x \,.\label{action} \ee One varies the fields $h_{\m \n}$ and the
coordinates $y^\m$ of the string (with the condition that it remains
attached to the magnetic source), but not the trajectories of the
sources, which are fixed because of the conservation laws
$\partial_\m T^{\m \n}=0$ and $\partial_\m \Theta^{\m \n}=0$. This
is a well known limitation of the linearized theory, present already
in the pure electric case.  To treat the sources as dynamical, one
needs to go beyond the linear theory.

If there is no magnetic source, the first term in the action
reduces to $$ S^{PF} = \frac{1}{16 \pi G} \int \frac{1}{4}\left( -
\partial_\l h_{\a \b} \partial^\l h^{\a \b} + 2 \partial_\l h^{\l \a}
\partial^\m h_{\m \a} -2 \partial^\l h \partial_\m h^{\m \l}
+ \partial_\l h \partial^\l h \right), $$ which is the Pauli-Fierz
action, whose variation with respect to $h_{\a \b}$ gives
$-\frac{1}{16 \pi G}$ times the linearized Einstein tensor $G^{\a
\b}$.  It is straightforward to verify that the variation of the
first term in the action with respect to $h_{\a \b}$ still gives
$-\frac{1}{16 \pi G}$ times the linearized Einstein tensor $G^{\a
\b}$ with correct $\Phi^{\m \n}_{\; \; \; \; \a}$ contributions even
in the presence of magnetic sources. So, the equations of motion
that follow from (\ref{action}) when one varies the gravitational
field are the Einstein equations (\ref{Einstein}).

Extremization with respect to the string coordinates does not bring
in new conditions provided that the Dirac string does not go through
an electric source (Dirac veto).

The action (\ref{action}) was obtained by using the analysis of
source-free linearized gravity in terms of two independent fields
given in \cite{Boulanger}, which enables one to go from the electric
to the magnetic formulations and vice-versa, by elimination of
magnetic or electric variables. As one knows how to introduce
electric sources in the electric formulation, through standard
minimal coupling, one can find how these sources appear in the
magnetic formulation by eliminating the electric variables and
keeping the magnetic potentials.  So, one can determine how to
introduce electric poles in the magnetic formulation, or, what is
equivalent, magnetic poles in the electric formulation.

\subsubsection{Gauge invariances}
{\it \small Diffeomorphism invariance} \vspace{.2cm}

The action (\ref{action}) is invariant under linearized
diffeomorphisms and under displacements of the Dirac string
(accompanied by appropriate transformations of the spin 2 field).
The easiest way to show this is to observe that the first term in
the action (\ref{action}) is invariant if one shifts $Y^{\m \n}_{\;
\; \; \; \a}$ according to \be Y^{\m \n}_{\; \; \; \; \a}
\rightarrow Y^{\m \n}_{\; \; \; \; \a} + \delta^\m_\a
\partial_\r z^{\n \r} - \delta^\n_\a \partial_\r z^{\m \r} +
\partial_\a z^{\m \n} \,,\label{gauge1}\ee where $z^{\m\n} = - z^{\n
\m}$ is arbitrary. This is most directly verified by noting that
under (\ref{gauge1}), the tensor $\bar{Y}^{\m \n}_{\; \; \; \; \a}$
defined in (\ref{ybar}) transforms simply as \be \bar{Y}^{\m \n}_{\;
\; \; \; \a} \rightarrow \bar{Y}^{\m \n}_{\; \; \; \; \a}  +
\partial_\a z^{\m \n} \label{gauge2}\ee and this leaves invariant
the first term in (\ref{action}) up to a total derivative. Note that
the Riemann tensor (\ref{curvature}) is strictly invariant.  The
transformation (\ref{gauge1}) can be conveniently rewritten as \be
Y^{\m \n}_{\; \; \; \; \a} \rightarrow Y^{\m \n}_{\; \; \; \; \a} +
\epsilon^{\m \n \r \s}
\partial_\r a_{\s \a} \,, \label{gauge3}\ee where $a_{\s \a} = - a_{
\a \s}$ is given by $ a_{\s \a} = \frac{1}{2} \epsilon_{\s \a \b
\g} z^{\b \g} $.

A (linearized) diffeomorphism \be h_{\m \n} \rightarrow h_{\m \n}
+ \partial_\m \xi_\n + \partial_\n \xi_\m \ee (with the string
coordinates unaffected) modifies $Y^{\m \n}_{\; \; \; \; \a}$ as
in (\ref{gauge3}) with $ a_{\s \a} =
\partial_\a \xi_\s - \partial_\s \xi_\a $ (note that the term
$\partial_\s \xi_\a$ in $a_{\s \a}$ does not contribute because
$\partial_{[\r}\partial_{\s]} \xi_\a = 0$).  Hence, the first term
in the action (\ref{action}) is invariant under diffeomorphisms. The
minimal coupling term is also invariant because the energy-momentum
tensor is conserved. It follows that the complete action
(\ref{action}) is invariant under diffeomorphisms. \vspace{2.3cm}

{\it \small Displacements of the Dirac string} \vspace{.2cm}

An arbitrary displacement of the Dirac string, \be y^\a (\tau, \s)
\rightarrow y^\a (\tau, \s) + \delta y^\a (\tau, \s)
\label{displ1}\ee also modifies $Y^{\m \n}_{\; \; \; \; \a}$ as in
(\ref{gauge3}) provided one transforms simultaneously the spin 2
field $h_{\m \n}$ appropriately. Indeed, under the displacement
(\ref{displ1}) of the Dirac string, the quantity $\Phi^{\m \n}_{\;
\; \; \; \a}$ changes as $ \Phi^{\m \n}_{\; \; \; \; \a} \rightarrow
\Phi^{\m \n}_{\; \; \; \; \a} + k^{\m \n}_{\; \; \; \; \a}$ where
$k^{\m \n}_{\; \; \; \; \a}$ can be computed from $\delta y^\a
(\tau, \s)$ through (\ref{exprforphi}) and has support on the old
and new string locations.  Its explicit expression will not be
needed. What will be needed is that it fulfills \be
\partial_\m k^{\m \n}_{\; \; \; \; \a} = 0 \,,\label{eqfork}\ee because
the magnetic energy-momentum tensor is not modified under a
displacement of the Dirac string. The field $Y^{\m \n}_{\; \; \; \;
\a}$ changes then as \be Y^{\m \n}_{\; \; \; \; \a} \rightarrow
Y^{\m \n}_{\; \; \; \; \a} + \epsilon^{\m \n \r \s} \partial_\r
\delta h_{\s \a} + k^{\m \n}_{\; \; \; \; \a} \label{eqforY}\ee
where $\delta h_{\s \a}$ is the sought after variation of $h_{\s
\a}$. By using (\ref{eqfork}), one may rewrite the last term in
(\ref{eqforY}) as $\partial_\r t^{\m \n \r}_{\; \; \; \; \; \; \a}$
for some $ t^{\m \n \r}_{\;  \; \; \; \; \; \a} =  t^{[\m \n
\r]}_{\; \; \; \; \; \; \; \a}$.  Again, we shall not need an
explicit expression for $t^{\m \n \r}_{\;  \; \; \; \; \; \a}$, but
only the fact that because $k^{\m \n}_{\; \; \; \; \a}$ has support
on the string locations, which do not go through the electric
sources (Dirac veto), one may choose $t^{\m \n \r}_{\; \; \; \; \;
\; \a}$ to vanish on the electric sources as well.  In fact, one may
take $t^{\m \n \r}_{\; \; \; \; \; \; \a}$ to be non-vanishing only
on a membrane supported by the string. Decomposing $t^{\m \n \r}_{\;
\; \; \; \; \; \a}$ as $ t^{\m \n \r}_{\;  \; \; \; \; \; \a} =
\epsilon^{\m \n \r \s}
 \left( s_{\s \a} + a_{\s \a} \right)$,
$s_{\s \a} = s_{(\s \a)}$, $a_{\s \a} = a_{[\s \a]}$ and taking
$h_{\s \a}$ to transform as $ h_{\s \a} \rightarrow h_{\s \a} -
s_{\s \a} $ one sees from (\ref{eqforY}) that the variation of
$Y^{\m \n}_{\; \; \; \; \a}$ takes indeed the form (\ref{gauge3}).
Thus the first term in the action is invariant. The minimal
coupling term is also invariant because the support of the
variation of the spin 2 field does not contain the electric
worldlines.

One can also observe that the variation $\d r_{\a \b \r \s}$
vanishes outside the original and displaced string locations. This
implies $\d h_{\a \b} = \partial_\a \xi _\b +
\partial_\b \xi_\a$ except on the location of both strings, where
$\xi_\a$ induces a delta function contribution on the string
(``singular gauge transformation"). The explicit expressions will
not be given here. \vspace{.3cm}

{\it \small Identities} \vspace{.2cm}

The identities which follow from the invariance (\ref{gauge1}), or
(\ref{gauge3}), of the first term $$ {\cal L} = \frac{1}{64 \pi
G}\left( \bar{Y}_{\a \b \g} \bar{Y}^{\a \g \b} -
\bar{Y}_\a \bar{Y}^\a \right) $$ in the action may be written as \be
\partial_\r \left(\frac{\partial {\cal L}}{\partial Y^{\a \b}_{\;
\; \; \; \g}}\right) \epsilon^{\a \b \r \s} =
\partial_\r \left(\frac{\partial {\cal L}}{\partial Y^{\a \b}_{\;
\; \; \; \s}}\right) \epsilon^{\a \b \r \g}\,.\ee They imply that
\beq - \frac{1}{16 \pi G} G^{\a \b} = \frac{\delta {\cal L}}{\delta
h_{\a \b}} &=& -\partial_\r \left(\frac{\partial {\cal L}}{\partial
Y^{\m \n}_{\; \; \; \; \b}}\right) \epsilon^{\m \n \r \a} \\ &=& -
\partial_\r \left(\frac{\partial {\cal L}}{\partial Y^{\m \n}_{\;
\; \; \; \a}}\right) \epsilon^{\m \n \r \b} \,,\eeq from which the
contracted Bianchi identities are easily seen to indeed hold.

The expression (\ref{curvature}) of the Riemann tensor in terms of
$\bar{Y}^{\m \n}_{\; \; \; \; \a}$ makes it clear that it is
invariant under (\ref{gauge2}) and thus, invariant under both
diffeomorphisms and displacements of the Dirac string.

\subsection{Quantization condition} Because of the gauge
invariances just described, the Dirac string is classically
unobservable.  In the Hamiltonian formalism, this translates itself
into the existence of first-class constraints expressing the momenta
conjugate to the string coordinates in terms of the remaining
variables. Demanding that the string remains unobservable in the
quantum theory imposes a quantization condition on the charges,
which we now derive. The argument follows closely that of Dirac in
the electromagnetic case \cite{Dirac:1948um}.

Working for simplicity in the gauge $y^0 = \l$ (which eliminates
$y^0$ as an independent variable), one finds the constraints \be
\label{just} \pi_m = - 32 \pi G \, N y'^n \, v_\g \frac{\partial
{\cal L}}{\partial Y^{mn}_{\; \; \; \; \g}} \,. \ee  The right hand
side of (\ref{just}) generates the change of the gravitational field
that accompanies the displacement of the Dirac string.

In the quantum theory, the wave functional $\psi$ must therefore
fulfill $$ \frac{\hbar}{i} \frac{\delta \psi}{\delta y^m(\s)} = - 32
\pi G \, N y'^n \, v_\g \frac{\partial {\cal L}}{\partial Y^{mn}_{\;
\; \; \; \g}} \psi \,.$$  We integrate this equation as in
\cite{Dirac:1948um}, along a path in the configuration space of the
string that encloses an electric source.  One finds that the
variation of the phase of the wave functional is given by \be \Delta
\Psi = - \frac{16 \pi G \, N \, v_\g}{\hbar} \int \frac{\partial
{\cal L}}{\partial Y^{mn}_{\; \; \; \; \g}} \left(\dot{y}^m y'^n -
\dot{y}^n y'^m \right) d \s d \l \,,\ee where the integral is taken
on the two-dimensional surface enclosing the electric source. Using
the Gauss theorem, this can be converted to a volume integral, $$
\Delta \Psi = - \frac{16 \pi G \, N \, v_\g}{\hbar} \int d^3 x \,
\e^{mnp}\,
\partial_p \left(\frac{\partial {\cal L}}{\partial Y^{mn}_{\; \; \;
\; \g}} \right) \,.$$ Because $ \e^{mnp}
\partial_p \left(\frac{\partial {\cal L}}{\partial Y^{mn}_{\; \;
\; \; \g}} \right) = \frac{\delta {\cal L}}{\delta h_{0 \g}} $, the
variation of the phase becomes, upon use of the constraint (initial
value) Einstein equations $G^{0 \g} = 8 \pi G T^{0 \g}$, $$ \Delta
\Psi = \frac{8 \pi G \, N \, v_\g}{\hbar} \int d^3 x T^{0 \g} =
\frac{8 \pi G \, N \, M \, v_\g u^\g}{\hbar}.$$ For the wave
functional to be single-valued, this should be a multiple of $2
\pi$.  This yields the quantization condition \be \frac{4 N M G v_\g
u^\g}{\hbar} = n , \; \; \; n \in \Bbb{Z} \,.\ee Introducing the
conserved charges $P^\g$, $Q^\g$ associated with the spin 2 theory
(electric and magnetic 4-momentum), this can be rewritten as \be
\frac{4  G P_\g Q^\g}{\hbar}  \in \Bbb{Z} \,.\ee It is to be
stressed that the quantization condition is not a condition on the
electric and magnetic masses, but rather, on the electric and
magnetic 4-momenta. In the rest frame of the magnetic source, the
quantization condition becomes \be \frac{4  G E N}{\hbar} \in
\Bbb{Z} \,,\ee where $E$ is the (electric) energy of the electric
mass. Thus, it is the energy which is quantized, not the mass.

We have taken above a pure electric source and a pure magnetic pole.
We could have taken dyons, one with charges $(P^\g, Q^\g)$, the
other with charges $(\bar{P}^\g, \bar{Q}^\g)$. Then the quantization
condition reads \be \frac{4 G \left(P_\g \bar{Q}^\g- \bar{P}_\g
Q^\g\right)}{\hbar} \equiv \frac{4 G
 \epsilon_{ab}Q^{a}_\g \bar{Q}^{b\g}}{\hbar}\in \Bbb{Z} \,,\ee
since the sources are pointlike (0-dyons). Here $Q^{a}_\g \equiv
(P_\g, Q_\g)$, $a,b = 1,2$ and $\epsilon_{ab}$ is the
$SO(2)$-invariant Levi-Civita tensor in the 2-dimensional space of
the charges.

\subsection{One-particle solutions}
\subsubsection{Electric mass} We consider a point
particle electric mass at rest at the origin of the coordinate
system. The only non-vanishing component of its electric energy
momentum tensor is $ T^{00}(x^0, \vec{x}) = M \delta^{(3)}(\vec{x})$
while $\Theta^{\m \n}$ vanishes. There is no Dirac string since
there is no magnetic mass. The metric generated by this source is
static. The linearized Einstein equations are well known to imply in
that case the linearized Schwarzschild solution, namely in polar
coordinates $$ h_{00} = \frac{2 GM}{r} = h_{rr}, \; \; \;
\hbox{other components vanish}\,,
$$ or in Cartesian coordinates $$ h_{00} = \frac{2 GM}{r}, \; \;
\; h_{ij} = \frac{2 GM}{r^3}x_i x_j, \; \; \; \hbox{other
components vanish.} $$ Indeed, one then finds \beq R_{0s0b} &=& M
\left(-\frac{3x_sx_b}{r^5} + \frac{\d_{sb}}{r^3} +
\frac{4 \pi}{3}\, \d_{sb} \, \d (\vec{x}) \right) \nonumber \\
R_{0sab} &=& 0  \; = \;
R_{ab0s}\,, \nonumber \\
R_{pqab} &=& \left(\d_{pa} \d_{qb} - \d_{pb} \d_{qa} \right)\left(
\frac{2M}{r^3} + \frac{8 \pi}{3} \d(\vec{x}) \right) \,,\nonumber
\\ && - 3 M \left( \d_{pa} \frac{x_b x_q}{r^5} - \d_{qa}
\frac{x_b x_p}{r^5} - \d_{pb} \frac{x_a x_q}{r^5} + \d_{qb}
\frac{x_a x_p}{r^5}\right) \,,\nonumber \eeq and thus $ R_{00} = 4
\pi G \, M \delta^{3}(\vec{x})$, $R_{ab} = 4 \pi G \, M \, \d_{ab}
\, \delta^{3}(\vec{x})$. The solution can be translated and boosted
to obtain a moving source at an arbitrary location.

\subsubsection{Magnetic mass} We now consider the dual solution,
that is, a point magnetic mass sitting at the origin. We have
$\Theta^{00}(x^0, \vec{x}) = N \delta^{(3)}(\vec{x})$ as the only
non-vanishing component of the magnetic energy-momentum tensor.
{}Furthermore, $T^{\m \n} = 0$. The solution is linearized Taub-NUT
\cite{Newman:1963yy}, with only magnetic mass, i.e., in polar
coordinates, $$ h_{0\varphi} = -2 N (1- \cos \theta), \; \; \;
\hbox{other components vanish.} $$ With this choice of
$h_{0\varphi}$ the string must be taken along the negative $z$-axis
in order to cancel the singularity at $\theta=\pi$. The tensor
$\Phi^{\a \b}_{\; \; \; \; \l}$ is given by $ \Phi^{0z}_{\; \; \; \;
0} = - 16 \pi N \theta(-z) \d(x) \d(y) $ (other components vanish).

One then finds the only non-vanishing components (in Cartesian
coordinates)
$$ \bar{Y}'^{0s}_{\; \; \; \; 0} = - 2 N \frac{x^s}{r^3}, \; \; \;
\; \bar{Y}'^{rs}_{\; \; \; \; c} = 2 N \left( \d^r_c \,
\frac{x^s}{r^3} - \d^s_c \, \frac{x^r}{r^3} \right) \,.$$ Here,
$\bar{Y}'^{\a \b}_{\; \; \; \; \g}$ differs from $\bar{Y}^{\a
\b}_{\; \; \; \; \g}$ by a gauge transformation (\ref{gauge2}) with
$z^{lm} = \epsilon^{lmp} h_{0p}$, $z^{0m} = 0$, and hence gives the
same curvature. Dealing with $\bar{Y}'^{\a \b}_{\; \; \; \; \g}$
rather than $\bar{Y}^{\a \b}_{\; \; \; \; \g}$ simplifies the
computations. It follows that the curvature is given by \beq
R_{0s0b} &=& 0, \; \; \; \; R_{lmab} = 0, \nonumber
\\R_{lm0b} &=& N \epsilon_{lms} \left(\frac{3 x_b x_s}{r^5} -
\frac{\d_{bs}}{r^3} - \frac{4 \pi}{3} \, \d_{bs}\, \d(\vec{x})
\right) , \nonumber \\ R_{0mab} &=& 2N
\epsilon_{abm}\left(\frac{1}{r^3} + \frac{4 \pi}{3} \d(\vec{x})
\right) - 3N \left( \epsilon_{mak} \frac{x_b x_k}{r^5} -
\epsilon_{mbk} \frac{x_a x_k}{r^5}\right)\,, \nonumber \eeq which
satisfies the equations of motion, $ R_{\a \b} = 0$ and $ R_{0ijk} +
R_{ij0k} + R_{j0ik} = 4 \pi N \epsilon_{ijk} \d(\vec{x}) = - 8 \pi
\epsilon_{0ij\l} \bar{\Theta}^\l_{\; k} $.

{}Finally, one easily checks that the linearized Riemann tensor of
linearized Taub-NUT is indeed dual to the linearized Riemann tensor
of linearized Schwarschild.  In that respect, the reason that it was
more convenient to work with $\bar{Y}'^{\a \b}_{\; \; \; \; \g}$
instead of $\bar{Y}^{\a \b}_{\; \; \; \; \g}$ above is that it is
$\bar{Y}'^{\a \b}_{mag \; \g}$ that is dual to $\bar{Y}^{\a
\b}_{Schw\; \g}$. While the curvatures are dual, the original
quantities $\bar{Y}^{\a \b}_{\; \; \; \; \g}$ are dual up to a gauge
transformation (\ref{gauge2}).

\section{Magnetic sources for bosonic higher spins}
\setcounter{equation}{0} \label{higherspins0}

We now indicate how to couple magnetic sources to spins greater than
two. The procedure parallels what we have just done for spin 2 but
the formulas are somewhat cumbersome because of the extra indices on
the fields and the extra trace conditions to be taken into account.
We shall therefore only give the general outline of the analysis in
the text and relegate the technical details to appendices.  The
formalism describing higher spin fields in the absence of magnetic
sources is recalled in Appendix \ref{appendixA}.

The spin s curvature $R_{ \m_1 \n_1 \m_2 \n_2 \cdots \m_s\n_s}$ is
the gauge invariant object in terms of which we shall first write
the equations of the theory.  Its index symmetry is described by
the Young tableau \be
\begin{picture}(85,15)(0,2)
\multiframe(0,11)(13.5,0){1}(10.5,10.5){\tiny{$\m_1$}}
\multiframe(0,0)(13.5,0){1}(10.5,10.5){\tiny{$\n_1$}}
\multiframe(31,11)(13.5,0){1}(10.5,10.5){\tiny{$\m_2$}}
\multiframe(31,0)(13.5,0){1}(10.5,10.5){\tiny{$\n_2$}}
\multiframe(42,11)(13.5,0){1}(19.5,10.5){$\cdots$}
\multiframe(42,0)(13.5,0){1}(19.5,10.5){$\cdots$}
\multiframe(62,11)(13.5,0){1}(10.5,10.5){\tiny{$\m_s$}}
\multiframe(62,0)(13.5,0){1}(10.5,10.5){\tiny{$\n_s$}}
\put(15,7.5){$\otimes$}
\end{picture}\,, \label{Young00}
\ee i.e., \be R_{ \m_1 \n_1 \cdots \m_i \n_i \cdots \m_s\n_s} = -
R_{ \m_1 \n_1 \cdots \n_i \m_i \cdots \m_s\n_s}, \; \; \; i = 1,
\cdots, s \ee and \be R_{ \m_1 \n_1 \cdots [\m_i \n_i
\m_{i+1}]\cdots \m_s\n_s} = 0 , \; \; \; i = 2, \cdots, s-1 \, .
\label{cycyclic}\ee Its dual, defined through
$$S_{\m_1 \n_1 \m_2 \n_2 \cdots \m_s\n_s} =-\frac{1}{2} \e_{\m_1\n_1
\r \s}R^{\r\s}_{~~\m_2 \n_2 \cdots \m_s\n_s} \, ,$$ has the same
symmetry structure. Note that, just as in the spin 2 case, this does
not define an irreducible representation of the linear group.  But,
also as in the spin 2 case, we shall find that only the irreducible
part described by \be
\begin{picture}(85,15)(0,2)
\multiframe(0,11)(13.5,0){1}(10.5,10.5){\tiny{$\m_1$}}
\multiframe(0,0)(13.5,0){1}(10.5,10.5){\tiny{$\n_1$}}
\multiframe(11,11)(13.5,0){1}(10.5,10.5){\tiny{$\m_2$}}
\multiframe(11,0)(13.5,0){1}(10.5,10.5){\tiny{$\n_2$}}
\multiframe(22,11)(13.5,0){1}(19.5,10.5){$\cdots$}
\multiframe(22,0)(13.5,0){1}(19.5,10.5){$\cdots$}
\multiframe(42,11)(13.5,0){1}(10.5,10.5){\tiny{$\m_s$}}
\multiframe(42,0)(13.5,0){1}(10.5,10.5){\tiny{$\n_s$}}
\end{picture}\, \label{Young000} \ee (i.e., fulfilling also
(\ref{cycyclic}) for $i=1$) corresponds to the independent degrees
of freedom (the rest being determined by the sources).

The electric and magnetic  energy-momentum tensors will be denoted
by $t_{\m_1 \m_2 \cdots \m_s} $ and $\theta_{\m_1 \m_2 \cdots
\m_s}$. They are conserved, i.e. divergence-free: $ \pa_\m
t^{\m\n_1 \cdots \n_{s-1}} =0$, $ \pa_\m \theta^{\m \n_1 \cdots
\n_{s-1}} =0$. Their double traceless parts are written $T_{\m_1
\m_2 \cdots \m_s} $ and $\Theta_{\m_1 \m_2 \cdots \m_s} $, and are
the tensors that actually couple to the spin s field.

\subsection{Electric and magnetic sources}

The equations in the presence of both electric and magnetic sources
are determined again by the requirements: (i) that they reduce to
the known equations with electric sources only when the magnetic
sources are absent, and (ii) that they be invariant under the
duality transformations that rotate the spin s curvature and its
dual, as well as the electric and magnetic sources.

Defining $\bar{\Theta}_{\m_1\m_2 \cdots \m_s}=\Theta_{\m_1\m_2
\cdots \m_s}- \frac{s}{4}
\eta_{(\m_1\m_2}\Theta^{\prime}_{\m_3\cdots \m_s)}\,,$ one finds
the following set of equations for the curvature: \beq R_{ \m_1
\n_1 \m_2 \n_2 \cdots \m_s\n_s} \eta^{\n_1\n_2}&=&\frac{1}{2}\
\bar{T}_{\m_1\m_2 [\m_3 [\cdots [\m_s,\n_s] \cdots ]\n_3]}\,, \label{CC1}\\
R_{ [\m_1 \n_1 \m_2] \n_2 \cdots \m_s\n_s}&=&\frac{1}{6} \
\e_{\m_1 \n_1\m_2
\r}\bar{\Theta}^{\r}_{~\n_2  [\m_3 [\cdots [\m_s,\n_s] \cdots ]\n_3]}\,,\label{CC2}\\
\partial_{[\a} R_{\m_1 \n_1] \m_2 \n_2 \cdots \m_s\n_s} &=&-\frac{1}{3}\
\e_{\a\m_1\n_1\r}\bar{\Theta}^{\r}_{~[\m_2  [\m_3 [\cdots
[\m_s,\n_s] \cdots ]\n_3] \n_2]}\,.\label{CC3}\eeq The first
equation is the analog of the Einstein equation (\ref{Einstein}),
the second is the analog of the modified cyclic identity
(\ref{cyclic}), while the third is the analog of the modified
Bianchi identity (\ref{BI}). It follows from these equations that
the dual curvature obeys similar equations, \beq S_{ \m_1 \n_1 \m_2
\n_2 \cdots \m_s\n_s} \eta^{\n_1\n_2}&=&\frac{1}{2}\
\bar{\Theta}_{\m_1\m_2 [\m_3 [\cdots [\m_s,\n_s] \cdots ]\n_3]} \,,\\
S_{ [\m_1 \n_1 \m_2] \n_2 \cdots \m_s\n_s}&=&-\frac{1}{6} \
\e_{\m_1 \n_1\m_2 \r}\bar{T}^{\r}_{~\n_2  [\m_3 [\cdots [\m_s,\n_s] \cdots ]\n_3]}\,,\\
\partial_{[\a} S_{\m_1 \n_1] \m_2 \n_2 \cdots \m_s\n_s}& =&\frac{1}{3}\
\e_{\a\m_1\n_1\r}\bar{T}^{\r}_{~[\m_2  [\m_3 [\cdots [\m_s,\n_s]
\cdots ]\n_3] \n_2]}\label{CC6}\, ,\eeq exhibiting manifest
duality symmetry.

\subsection{Decomposition of the curvature tensor}

As in the spin 2 case, the curvature tensor can be expressed in
terms of a completely symmetric potential $h_{\m_1 \cdots \m_s}$
and of a tensor $\Phi^{\r\s}_{~~~\m_1 \cdots \m_{s-1}}$ fixed by
the magnetic energy-momentum tensor, so that the cyclic and
Bianchi identities do indeed become identities.

Let $\Phi^{\r\s}_{~~~\m_1 \cdots \m_{s-1}}$ be such that \be\pa_\r
\Phi^{\r\s}_{~~~\m_1 \cdots \m_{s-1}}=\theta^\s_{~\m_1 \cdots
\m_{s-1}}\,, \label{phis}\ee and let $\hat{\Phi}^{\r\s}_{~~~\m_1
\cdots \m_{s-1}}$ be the part of $\Phi^{\r\s}_{~~~\m_1 \cdots
\m_{s-1}}$ that is traceless in the indices $\m_1 \cdots \m_{s-1}$.
For computations, it is useful to note that $ \pa_\r
\hat{\Phi}^{\r\s}_{~~~\m_1 \cdots \m_{s-1}}=\Theta^\s_{~\m_1 \cdots
\m_{s-1}}-\frac{(s-2)}{4} \eta_{(\m_1\m_2}\Theta'^{\s}_{~\m_3\cdots
\m_{s-1})}\,. $ The expression of the tensor $\Phi^{\r\s}_{~~~\m_1
\cdots \m_{s-1}}$ in terms of the Dirac string is given below. The
appropriate expression of the curvature tensor in terms of the spin
s field and the Dirac string contribution is: \be R_{ \m_1 \n_1 \m_2
\n_2 \cdots \m_s\n_s} =-\frac{1}{2}\ \e_{\m_1\n_1 \r\s}
\bar{Y}^{\r\s}_{~~~[\m_2 [\m_3 [\cdots [\m_s,\n_s] \cdots ]\n_3]
\n_2]}\,, \ee where \beq\bar{Y}^{\r\s}_{~~~~\m_1 \cdots
\m_{s-1}}&=&Y^{\r\s}_{~~~~\m_1 \cdots
\m_{s-1}}+\frac{2(s-1)}{s}\d^{[\r}_{(\m_1}Y^{\s]\t}_{~~~~\m_2
\cdots \m_{s-1})\t}\,,\label{YYY1}\\
Y^{\r\s}_{~~~~\m_1 \cdots \m_{s-1}}&=&\pa_\t
X^{\r\s\t}_{~~~\m_1 \cdots \m_{s-1}}+
\hat{\Phi}^{\r\s}_{~~~\m_1 \cdots \m_{s-1}}\,,\label{YYY2}\\
X^{\r\s\t}_{~~~~\m_1 \cdots \m_{s-1}}&=&\e^{\r\s\t\l}h_{\l\m_1
\cdots \m_{s-1}} -\frac{3(s-1)(s-2)}{2s}\eta_{\a
(\m_1}\d_{\m_2}^{[\r}\e^{\s\t]\a\b}h_{\m_3\cdots
\m_{s-1})\b}^{\prime}\,.\label{YYY3} \eeq The split of
$Y^{\r\s}_{~~~~\m_1 \cdots \m_{s-1}}$ into an $X$-part and a
$\Phi$-part defines a split of the Riemann tensor analogous to the
split (\ref{split0}) introduced for spin 2.  The Dirac string
contribution ($\Phi$-term) removes the magnetic terms violating the
standard cyclic and Bianchi identities, leaving one with a tensor
$r_{ \m_1 \n_1 \m_2 \n_2 \cdots \m_s\n_s}$ that fulfills $$r_{ [\m_1
\n_1 \m_2] \n_2 \cdots \m_s\n_s}=0, \;\;\;
\partial_{[\a} r_{\m_1 \n_1] \m_2 \n_2 \cdots \m_s\n_s} =0 \,,$$ and
thus derives from a symmetric potential (the spin s field $h_{\m_1
\cdots \m_s}$) as \be r_{\m_1 \n_1 \m_2 \n_2 \cdots \m_s \n_s} =-2\
h_{[\m_1[ \m_2 \cdots [\m_s,\n_s]\cdots \n_2]\n_1]} \label{DefR0}\ee
(see Appendix \ref{appendixA}).  The $X$-term in the curvature is a
rewriting of (\ref{DefR0}) that is convenient for the subsequent
analysis. The potential $h_{\m_1 \cdots \m_s}$ is determined from
the curvature up to a gauge transformation with unconstrained trace.
The fact that only $\hat{\Phi}^{\r\s}_{~~~\m_1 \cdots \m_{s-1}}$
appears in the curvature and not $\Phi^{\r\s}_{~~~\m_1 \cdots
\m_{s-1}}$ is a hint that only the double traceless part
$\Theta_{\m_1 \cdots \m_{s}}$ of the magnetic energy-momentum tensor
plays a physical role.

\subsection{Equations of motion for the spin s field}
In terms of the potential, the remaining equation (\ref{CC1}) is of
order s.  In the sourceless case, one replaces it by a second order
equation written first by Fronsdal \cite{Fronsdal:1978rb}. This can
be done also in the presence of both electric and magnetic sources
by following the procedure described in
\cite{Francia:2002pt,Bekaert:2003az}. The crucial observation is
that the curvature is related as in (\ref{FR}), namely, \be R_{ \m_1
\n_1 \m_2 \n_2 \cdots \m_s\n_s}
\eta^{\n_1\n_2}=-\frac{1}{2}F_{\m_1\m_2 [\m_3 [\cdots [\m_s,\n_s]
\cdots ]\n_3]}\,,\label{FR0}\ee to the generalized Fronsdal tensor
given by \be F_{\g_1\cdots\g_s}=-\frac{1}{2}
\e_{\g_1\m\n\l}\Big(\pa^\l \bar{Y}^{\m\n}_{~~~\g_2\cdots\g_s}
-(s-1)\
\pa_{(\g_2}\bar{Y}^{\m\n\l}_{~~~~~\g_3\cdots\g_s)}\Big)\,,\label{fronsgen}
\ee so that (\ref{CC1}) is equivalent to $ F_{\m_1\m_2 [\m_3 [\cdots
[\m_s,\n_s] \cdots ]\n_3]} +\bar{T}_{\m_1\m_2 [\m_3 [\cdots
[\m_s,\n_s] \cdots ]\n_3]} = 0 $.  This implies $F_{\m_1\m_2 \m_3
\cdots \m_s}+\bar{T}_{\m_1\m_2 \m_3 \cdots \m_s} = \pa_{(\m_1
\m_2\m_3}\Lambda_{\m_4\cdots \m_s)}$ for some $\Lambda_{\m_4\cdots
\m_s}$ \cite{Olver}.  By making a gauge transformation on the spin s
field, one can set the right-hand side of this relation equal to
zero (see Appendix \ref{appendixA}), obtaining the field equation
\be \label{poppast}F_{\m_1\m_2 \m_3 \cdots \m_s}+\bar{T}_{\m_1\m_2
\m_3 \cdots \m_s} =0\,,\ee which fixes the trace of the gauge
parameter. When s = 3 this is the end of the story.

For s $\geq 4$ additional restrictions are necessary, namely, we
shall demand that the gauge transformation that brings the field
equation to the form (\ref{poppast}) eliminates at the same time the
double trace of the field $h_{\m_1 \cdots \m_s}$ (see
\cite{Francia:2002pt} for a discussion).

In terms of the generalized Einstein tensor defined as in
(\ref{eomsp}), i.e., \be G_{\m_1 \m_2 \cdots \m_s} =F_{\m_1 \m_2
\cdots \m_s} -\frac{s(s-1)}{4} \eta_{(\m_1\m_2}F_{\m_3\cdots
\m_s)\r}^{\hspace{1.2cm} \r} ,\label{EinsteinS0} \ee the equations
become \be G_{\m_1\m_2 \m_3 \cdots \m_s}+T_{\m_1\m_2 \m_3 \cdots
\m_s} =0.\label{EinHSEq0}\ee

We shall thus adopt (\ref{EinHSEq0}), with the Einstein tensor,
Fronsdal tensor and $Y$-tensor defined as in (\ref{EinsteinS0}),
(\ref{fronsgen}) and (\ref{YYY1}), respectively,  as the equations
of motion for a double traceless spin s field $h_{\m_1 \cdots
\m_s}$. These equations imply (\ref{CC1}) through (\ref{CC6}) and
define the theory in the presence of both electric and magnetic
sources. It is these equations that we shall derive from a
variational principle. 

\subsection{Point particles sources - Dirac string}

For point sources, the tensors that couple to the spin s field read
$$ t^{\m\n_1 \cdots \n_{s-1}} = M u^{\n_1} \cdots u^{\n_{s-1}}\int
d\l \delta^{(4)}(x-z(\l)) \dot{z}^\m =M \frac{u^\m u^{\n_1} \cdots
u^{\n_{s-1}}}{u^0} \delta^{(3)}(\vec{x} - \vec{z}(x^0))$$ and $$
 \theta^{\m \n_1 \cdots \n_{s-1}} =
N v^{\n_1} \cdots v^{\n_{s-1}} \int d\l \delta^{(4)}(x-\bar{z}(\l))
\dot{\bar{z}}^\m \\
=N \frac{v^\m v^{\n_1} \cdots v^{\n_{s-1}}}{v^0}
\delta^{(3)}(\vec{x} - \vec{\bar{z}}(x^0))\,.$$ One can check that
they are indeed conserved.

A tensor $\Phi^{\a \b}_{\; \; \; \; \g_1 \cdots \g_{s-1}}$ that
satisfies (\ref{phis}) can again be constructed by attaching a
Dirac string $y^\m(\l, \s)$ to the magnetic source, $y^\m(\l,
0)=\bar{z}^\m(\l)$. One has $$ \Phi^{\a \b}_{\; \; \; \; \g_1
\cdots \g_{s-1}} =  N v_{\g_1} \cdots v_{\g_{s-1}} \int d \l d \s
(y'^\a \dot{y}^\b  - \dot{y}^\a y'^\b) \delta^{(4)}(x-y(\l, \s))
\,.$$

One can compute explicitly the conserved charges associated with
asymptotic symmetries for electric point sources (see Appendix
\ref{appendixA}). Using the equations of motion, they read $$
P^{\m_1 \cdots \m_{s-1}}=M f^{\m_1 \cdots \m_{s-1}}(u)\,,$$ where
$f^{\m_1 \cdots \m_{s-1}}(u)$ is the traceless part of $u^{\m_1}
\cdots u^{\m_{s-1}}$. One then finds
$$ f^{\m_1 \cdots \m_{s-1}}(u)= \sum_l \a_l \ \eta^{(\m_1\m_2}
\cdots \eta^{\m_{2l-1} \m_{2l}}  u^{\m_{2l+1}} \cdots
u^{\m_{s-1})}\vert u\vert^{2l} \,,$$ where the sum goes over
$l=0,1,\cdots$ such that $2l\leq s-1 $, $\a_0=1$ and
$\a_{l+1}=-\frac{(s-1-2l)(s-2-2l)}{4(l+1)(s-1-l)}\ \a_l$ .

The dual magnetic charges
$$ Q^{\m_1 \cdots \m_{s-1}}=N f^{\m_1 \cdots \m_{s-1}}(v)\,$$ are
also conserved.

\subsection{Variational Principle}
The second-order equations of motion $
G_{\g_1\cdots\g_s}+\ T_{\g_1\cdots\g_s}=0$ equivalent to
(\ref{CC1}), are the Euler-Lagrange derivatives with respect to
$h^{\g_1\cdots\g_s}$ of the action \beq S[h_{\m_1 \cdots \m_s}(x),
y^\m(\l,\s)]=\int d^4x \ ( {\cal L} +\ h_{\m_1 \cdots
\m_s} t^{\m_1 \cdots \m_s} )\,,\label{actions} \eeq where \beq {\cal
L}=-\frac{(s-1)}{2}Y_{\m\n \a_1 \cdots \a_{s-1}}
\Big[-Y^{\m\a_1 \n\a_2 \cdots \a_{s-1}}+\frac{(s-2)}{2(s-1)}
Y^{\m\n \a_1 \cdots \a_{s-1}}\hspace{2cm}\nonumber \\
+\frac{(s-3)}{s}\eta^{\m\a_1}Y^{\n\r \a_2 \cdots
\a_{s-1}}_{\hspace{1.6cm}\r}-\frac{(s-2)}{s}\eta^{\m\a_1}Y^{\a_2\r
\n\a_3 \cdots \a_{s-1}}_{\hspace{2cm}\r}\Big]\,.\nonumber \eeq
Indeed, one can check that $\frac{\d {\cal L}}{\d
h^{\g_1\cdots\g_s}}=G_{\g_1\cdots\g_s}\,.$ As in the spin 2 case,
the trajectories of the electric and magnetic sources are kept
fixed, i.e., the sources are not dynamical. The magnetic coupling in
the action was obtained by introducing the familiar minimal electric
coupling in the ``parent action" of \cite{Boulanger}, which contains
two potentials, and determining what it becomes in the dual
formulation.

The action (\ref{actions}) can be verified to be invariant under the
gauge symmetries (\ref{gauges}) of the spin s field as well as under
displacements of the Dirac string (accompanied by an appropriate
redefinition of $h_{\m_1 \cdots \m_s}$).  This is done in Appendix
\ref{appendixGI}.

\subsection{Quantization condition}

As for spin 2, the unobservability of the Dirac string in the
quantum theory leads to a quantization condition.  The computation
proceeds exactly as in the spin 2 case and is reproduced in Appendix
\ref{QCHS}. One finds explicitly the quantization condition \be
\frac{MN  }{2 \pi \hbar} \, f_{\g_1 \cdots \g_{s-1}}(v)f^{\g_1
\cdots \g_{s-1}}(u)= n , \; \; \; n \in \Bbb{Z} \,.\ee In terms of
the conserved charges $P^{\g_1 \cdots \g_{s-1}}$ associated with the
asymptotic symmetries and their dual charges $Q^{\g_1 \cdots
\g_{s-1}}$, the quantization condition can be rewritten as \be
\frac{1 }{2 \pi \hbar} \, Q_{\g_1 \cdots \g_{s-1}}(v)P^{\g_1 \cdots
\g_{s-1}}(u) \in \Bbb{Z} \,.\ee

\section{Beyond the Linear Theory for Spin Two}
\label{sect4}
We have seen that magnetic sources can be introduced for linearized
gravity and linearized higher-spin theories, and that an appropriate
generalization of the Dirac quantization condition on the sources
must hold. However in the linear theory the treatment is already
incomplete since the sources must be external.  In the full non
linear theory even the introduction of external sources is not
possible.  For spin 2 the difficulty stems from the fact that the
source must be covariantly conserved and for spins $\geq 2$ 
the formulation of the non linear theory is still incomplete.

Nevertheless, we shall address the issue of duality in the spin 2
case, for which the non linear theory in the absence of sources is
the vacuum Einstein theory of gravitation. This is the ``electric''
formulation.  Electric sources are coupled through their standard
energy-momentum momentum tensor.  We do not know whether magnetic
sources should appear as independent fundamental degrees of freedom
(the complete action with these degrees of freedom included is
unknown and would presumably be non local, as the results of
\cite{Deser:2005sz} suggest) or whether they should
appear as solitons somewhat like in Yang-Mills theory
\cite{Montonen:1977sn}.

Whatever the answer, there are indications that duality invariance
and quantization conditions are valid beyond the flat space, linear
regime studied above. One indication is given by dimensional
reduction of the full Einstein theory, which reveals the existence
of ``hidden symmetries'' that include duality \cite{Geroch}. Another
indication that non linear gravity enjoys duality invariance is
given by the existence of the Taub-NUT solution
\cite{Newman:1963yy}, which is an exact solution of the vacuum
Einstein theory describing a gravitational dyon. The quantization
condition on the energy of a particle moving in the Taub-NUT
geometry is a well known result which has been discussed by many
authors \cite{Zeeetal} and which can be viewed as a consequence of
the existence of closed timelike lines \cite{Misner1}.

Because of the importance of the Taub-NUT metric, we shall devote
the rest of the paper to a further investigation of its
properties. {}First, we shall rederive the quantization condition
along new lines, from the quantization of the angular momentum
stored in the gravitational field.  Next, we compute the
Poincar\'e charges for the Taub-NUT metric, which is
asymptotically flat at spatial infinity in the sense of \cite{BR}.

\setcounter{equation}{0}
\subsection{The gravitating magnetic pole}

The Taub-NUT metric is given by :\be \label{metric1} ds^2 = - V(r)[
dt + 2N(k-\cos\theta) \, d\phi ]^2 + V(r)^{-1} dr^2 +
(r^2+N^2)(d\theta^2 + \sin^2 \theta\, d\phi^2) \,, \ee with \be V(r)
= 1 - \frac{2(N^2+Mr) }{(r^2 + N^2)} = \frac{r^2 - 2Mr -N^2}{r^2 +
N^2} \,, \ee where $N$ and $M$ are the magnetic and electric masses
as follows from the asymptotic analysis of the metric and our
discussion of the linear theory. A pure magnetic mass has $M=0$. The
number $k$ can be changed according to \be \label{gauge2'} k
\rightarrow k' = k - \alpha \ee by performing a $t$ coordinate
transformation \be \label{gauge1'} t \rightarrow t' = t + 2N \alpha
\phi\,. \ee

The metric (\ref{metric1}) is singular on the $z$-axis. This
singularity is most easily seen by calculating $|\nabla t|^2$ and
can be interpreted as a singularity of the metric or of the $t$
coordinate. It is known in the literature as the Dirac-Misner string
singularity, due to its analogy with the Maxwell case. Its location
depends on the value of $k$: for $k=1$ the singularity is at
$\theta=\pi$ and for $k=-1$ it is at $\theta=0$. For all other
values of $k$ both string singularities exist. The choice $k= 0$
makes the North and South poles play a symmetrical role. In his
paper \cite{Misner1}, Misner showed that the singularity is a
coordinate singularity and that the metric describes a non-singular
manifold provided that the time coordinate $t$ is taken to be
periodic with period $8\pi N$.

\subsection{Spatial rotations and quantization condition}

Even though the metric contains a $dt d\phi$ term, it is spherically
symmetric.  However, the rotation group acts on spacetime in an
unconventional way \cite{Misner1} and the rotation Killing vectors
differ from those of flat space by extra terms. One can understand
the origin of these extra terms by comparing the Taub-NUT solution
with the standard electromagnetic magnetic monopole solution and
recalling that for stationary metrics, the mixed time-space metric
components are naturally interpreted as the component of an
electromagnetic vector potential (``gravitomagnetism"): $g_{0 i}
\sim A_i$.  In this spirit, the above metric component $g_{0 i}$
would correspond to a monopole potential \be \label{pot2} A_\phi =
-2N (k-\cos\theta) \ee with magnetic charge $N$, as has been
observed by many authors.

To understand the form of the Killing vectors, we start with the
generator of rotations around the $z$-axis.  It is recalled in
Subsection \ref{rots} that in the electromagnetic case, rotations of
the electromagnetic potential of a magnetic pole must be accompanied
by gauge transformations. When the $g_{0i}$ metric components are
interpreted as the components of a vector potential, the
electromagnetic gauge transformations lift to diffeomorphisms along
the time direction. We therefore expect the gauge parameter which
accompanies a rotation to lift to a component along $\6/\6 t$.

This expectation turns out to be correct.  The metric
(\ref{metric1}) has four Killing vectors, given in \cite{Misner1}
for $k=1$, which we display as \bea \label{killings2} \xi_t & = &
\frac{\6}{\6 t}\,, \nn \xi_x & = & -\sin\phi \, \frac{\6}{\6 \theta}
- \cos\phi \cot\theta \, \frac{\6}{\6 \phi} + \left( 2Nk \cos\phi
\cot\theta -2N \frac{\cos\phi}{\sin\theta}\right)
\frac{\partial}{\partial t} \,, \nn \xi_y & = & \cos\phi \,
\frac{\6}{\6 \theta} - \sin\phi \cot\theta \, \frac{\6}{\6 \phi} +
\left( 2Nk \sin\phi \cot\theta - 2N \frac{\sin\phi}{\sin\theta}
\right)\frac{\partial}{\partial t} \,, \nn \xi_z & = & \frac{\6}{\6
\phi} - 2 N k\frac{\partial}{\partial t} \,. \eea These satisfy the
commutation relations \be [\xi_a,\xi_b] = - \epsilon_{abc} \xi_c\,,
\qquad [\xi_a,\xi_t] = 0\,, \ee where $a,b,c=x,y,z$ and
$\epsilon_{xyz}=1$, which constitute the standard su(2)$\times$u(1)
Lie algebra. For $N = 0$, which is the Schwarzschild case, the three
Killing vectors $\xi_a$ generate spacelike two-spheres.  In the case
of Taub-NUT, although the algebra is the same, the action of the
group on the manifold is, as pointed out by Misner, different. In
fact, the $\xi_a$ which satisfy the su(2) algebra now generate the
$r=$constant three-spheres.  These three-spheres have a Lorentzian
metric.

The three rotation Killing vectors $\xi_a$ are invariant under
reflections with respect to the origin ($t\rightarrow -t$,
$\theta\rightarrow \pi - \theta$, $\phi\rightarrow\phi + \pi$,
$N\rightarrow -N$ and $k\rightarrow -k$). This should be the case
because being the generators of rotations they are pseudo-vectors.

One also observes that $\xi_z$ has a component along
$\partial/\partial t$ proportional to the magnetic mass.  We find
this fact very appealing, as it leads to a different derivation of
the quantization condition as we now show.

Consider an electric mass following a geodesic in the Taub-NUT
spacetime, with four-momentum $p_{\m} = m u_\mu$.  There exists a
conserved charge associated with every Killing vector field, and in
particular a charge $J_Z$ associated with $\xi_z$: \be J_Z=m
\xi_z^\mu u_\mu = p_{\phi} - 2 N m u_0 k\,. \ee We see that just as
in the electromagnetic case, the angular momentum about the z-axis
has an extra piece besides the standard $p_\phi$. This extra piece
comes from the angular momentum in the field, which varies as the
particle moves and which must be taken into account in the
conservation law.  In fact, the extra term coincides with the
angular momentum in the field when $k=1$ and the particle is on the
positive $z$-axis (see Appendix \ref{appendixB}). Thus the angular
momentum in the field is equal to $2N m u_0$. Requiring this angular
momentum to be quantized in multiples of $\hbar/2$\ yields the
quantization condition \be 4 N m u_0 \in \hbar \Bbb{Z} \,.
\label{IV9}\ee Note that when the particle is on the negative
$z$-axis, the angular momentum in the field coincides with the extra
term in $J_Z$ when $k= -1$ and therefore changes sign.

Formula (\ref{IV9}) agrees with the condition that comes from
periodicity in time of the wave function.  Suppose indeed that the
electric mass is described by a wavefunction $\psi$. The time
dependence of $\psi$ is given by \be \psi \propto
e^{-\frac{iEt}{\hbar}} \ee where $E=mu_0$ is the energy of the
particle. Recalling that time is periodic with period $\Delta t =
8\pi N$, and requiring single valuedness of the wave function, we
obtain $E \Delta t = 2n \pi \hbar$ which implies, yet again, that
\be 4 N m u_0 \in \hbar \Bbb{Z} \,. \ee Single-valuedness is
required because the closed timelike curves are contractible.

At this point we may resort to the concept of duality invariance to
argue that the result should also hold if we were considering a test
magnetic mass in the gravitational field of an electric mass, or in
fact both. Thus we are led to the conclusion that the quantization
conditions derived in the linear theory \be \frac{4 G \left(P_\g
\bar{Q}^\g- \bar{P}_\g Q^\g\right)}{\hbar} \equiv \frac{4 G
\epsilon_{ab}Q^{a}_\g \bar{Q}^{b\g}}{\hbar}\in Z \ee also holds in
the full non linear theory.  The scalar product which appears here
is the Minkowskian scalar product of the asymptotic charges
$Q^{a}_\g$ and $\bar{Q}^{b\g}$.  The antisymmetry of the
quantization condition for the exchange of the dyons is confirmed by
the change of sign of the angular momentum stored in the field which
was observed above.

\subsection{Taub-NUT is asymptotically flat at spatial infinity}

The introduction of the Minkowskian scalar product is legitimate
because the Taub-NUT metric (\ref{metric1}) is asymptotically flat
at spatial infinity in the sense of \cite{BR}.  In polar
coordinates, the conditions of \cite{BR}, including their parity
conditions, read $h_{rr} = O(r^{-1})$, $h_{r \theta} = O(1)$, $h_{r
\phi} = O(1)$, $h_{\theta \theta} = O(r)$, $h_{\theta \phi} = O(r)$,
$h_{\phi \phi} = O(r)$ and the leading terms (of the order just
written) in $h_{rr}$, $h_{r\phi}$, $h_{\theta \theta}$, $h_{\phi
\phi}$ should be even under the inversion $\theta \rightarrow \pi -
\theta$ and $\phi \rightarrow \phi + \pi$, while the leading terms
in $h_{r\theta}$, $h_{\theta \phi}$ should be odd. Here, $h_{ij}$ is
the deviation from the flat metric also written in polar
coordinates. Similarly, the conjugate momenta should fulfill
$\pi^{rr} = O(1)$, $\pi^{r \theta} = O(r^{-1})$, $\pi^{r \phi} =
O(r^{-1})$, $\pi^{\theta \theta} = O(r^{-2})$, $\pi^{\theta \phi} =
O(r^{-2})$, $\pi^{\phi \phi} = O(r^{-2})$, with the parity
conditions that the leading terms in $\pi^{rr}$, $\pi^{r \phi}$,
$\pi^{\theta \theta}$ and $\pi^{\phi \phi}$ are odd, while those in
$\pi^{r \theta}$ and $\pi^{\theta \phi}$ are even.

It is easy to see that all these conditions are fulfilled in the
coordinate system with $k=0$, where the Taub-NUT metric reads \be
\label{metric1'} ds^2 = - V(r)[ dt - 2N\cos\theta \, d\phi ]^2 +
V(r)^{-1} dr^2 + (r^2+N^2)(d\theta^2 + \sin^2 \theta\, d\phi^2) \,.
\ee  The perturbation $h_{ij}$ is not smooth as a tensor on the
sphere because of the string singularities at $\theta =0, \pi$ where
$h_{\phi \phi}$ does not vanish, but its behavior as a function of
$r$ is nevertherless compatible with the required asymptotic
fall-off. This guarantees finiteness of all the surface integrals
for the Poincar\'e charges because the parity conditions are in
particular fulfilled.  Coordinate systems with $k \not=0$ fail to
fulfill the parity conditions and for this reason, are not amenable
to the treatment of \cite{BR}.

This notion of asymptotic flatness appears to be all that is needed
for the action principle and the identification of conserved
charges.  It does not, and need not, eliminate, of course, the
geometrical difficulties pointed out in \cite{Misner1} in regard,
for example, to the probable impossibility of writing the metric at
large distances as a smooth perturbation of Minkowski space.

Note that it is not necessary to fix the magnetic mass $N$ at
infinity. Competing histories in the variational principle may have
different $N$'s.  This treats $M$ and $N$ on an equal footing,
preserving duality \footnote{We have been kindly informed by Donald
Marolf that he can also vary the magnetic mass $N$ in a covariant
formalism that he has developed.}. It is at first sight disturbing
that the Killing vectors (\ref{killings2}) depend on $N$ because it
would seem that the asymptotic symmetry group depends on the
magnetic mass which as we just noted can be varied.  There is
however no difficulty because the surface integrals associated with
the $N$ dependent terms in the killing vectors vanish on account of
the parity conditions. Indeed as it was observed above since $N$ is
a pseudo-scalar the angular dependence of the factor that multiplies
it in the rotation Killing vectors is also odd under inversions as
required by the parity conditions.

If one computes the Poincar\'e charges for the metric
(\ref{metric1'}) using the formulas of \cite{BR}, one finds \be P^0
= M, \; \; \; P^i = 0, \; \; \; J_{ij} = 0, \; \; \; J_{0i} = 0
\,.\ee The parameter $N$ does not contribute to the ``electric"
Poincar\'e charges.  Observe in particular that the angular momentum
is zero for the Taub-NUT solution, indicating that the solution may
be thought of as describing a dyon with the two charges on top of
one another. If the magnetic mass and electric mass are separated,
there is a net angular momentum and the metric belongs to the
Kerr-Newman-Taub-NUT \cite{DemNew} family, which describes metrics
with angular momentum and electromagnetic charges. These metrics
fulfill also the boundary conditions of \cite{BR}. Other asymptotic
analysis of the Taub-NUT solution (at null infinity) may be found in
\cite{RSenAsh}.

\section{Conclusions}
\setcounter{equation}{0}

In this paper we have extended to all integer spins the action for
magnetic poles introduced by Dirac.  The construction has two
limitations: (i) it deals only with the linearized field theory and
(ii) the sources must be external.  The presumption is however
strong that, at least in the spin 2 case, that is, the theory of
gravitation, monopoles exist and the corresponding Dirac
quantization condition holds.  Arguments in this direction are
advanced based on properties of Taub-NUT space.  The analogy with
electromagnetism is emphasized throughout.

\vspace{1truecm}

\acknowledgments

MH is grateful to KITP for hospitality while the article was being
prepared. This work was funded by an institutional grant to CECS of
the Millennium Science Initiative, Chile, and Fundaci\'on Andes, and
also benefits from the generous support to CECS by Empresas CMPC.
The work of SC and MH is partially supported  by IISN - Belgium
(convention 4.4505.86), by the ``Interuniversity Attraction
Monopoles Programme -- Belgian Science Policy " and by the European
Commission programme MRTN-CT-2004-005104, in which they are
associated to V.U. Brussel. RP gratefully acknowledges support from
FONDECYT grant 3050086. This research was also supported in part by
the National Science Foundation under Grant No. PHY99-07949.
SC is Aspirant du FNRS (Belgium).

\newpage

\begin{appendix}
\section{Higher spin gauge fields in the absence of magnetic
sources} \label{appendixA}\setcounter{equation}{0}

We recall in this appendix the free theory of bosonic higher spin
fields.

\subsection{Spin s field and gauge invariance}

In the absence of sources, a massless bosonic spin $s$ field is
described by a gauge potential which is a totally symmetric tensor
$h_{\m_1 \m_2 \cdots \m_s}$ subject to the ``double-tracelessness
condition" \cite{Fronsdal:1978rb}, $ h_{\m_1 \m_2 \cdots \m_s} =
h_{(\m_1 \m_2 \cdots \m_s)}, \; \; \; h_{\m_1 \m_2 \m_3 \m_4 \cdots
\m_s} \eta^{\m_1 \m_2} \eta^{\m_3 \m_4} = 0 $.  The gauge
transformation reads \be h_{\m_1 \m_2 \cdots \m_s} \rightarrow
h_{\m_1 \m_2 \cdots \m_s} +
\partial_{(\m_1}\xi_{\m_2 \cdots \m_s)} \,,\label{gauges}\ee where the gauge
parameter $\xi_{\m_2 \cdots \m_s}$ is traceless,  $ \xi_{\m_2 \m_3
\cdots \m_s} \eta^{\m_2 \m_3}= 0 $.  The trace condition on the
gauge parameter appears for spins $\geq 3$, while the double
tracelessness condition on the field appears for spins $\geq 4$.

{}From the field $h_{\m_1 \m_2 \cdots \m_s}$, one can construct a
curvature $R_{\m_1 \n_1 \m_2 \n_2 \cdots \m_s \n_s}$ that contains
$s$ derivatives of the field and that is gauge invariant under the
transformations (\ref{gauges}) even if the gauge parameter is not
traceless, \be R_{\m_1 \n_1 \m_2 \n_2 \cdots \m_s \n_s} =-2\
h_{[\m_1[ \m_2 \cdots [\m_s,\n_s]\cdots \n_2]\n_1]}
\,,\label{DefR}\ee where one antisymmetrizes over $\m_k$ and $\n_k$
for each $k$. This is the analog of the Riemann tensor of the spin 2
case.  The curvature $R_{\m_1 \n_1 \m_2 \n_2 \cdots \m_s \n_s}$ has
the symmetry characterized by the Young tableau \be
\begin{picture}(85,15)(0,2)
\multiframe(0,11)(13.5,0){1}(10.5,10.5){\tiny{$\m_1$}}
\multiframe(0,0)(13.5,0){1}(10.5,10.5){\tiny{$\n_1$}}
\multiframe(11,11)(13.5,0){1}(10.5,10.5){\tiny{$\m_2$}}
\multiframe(11,0)(13.5,0){1}(10.5,10.5){\tiny{$\n_2$}}
\multiframe(22,11)(13.5,0){1}(19.5,10.5){$\cdots$}
\multiframe(22,0)(13.5,0){1}(19.5,10.5){$\cdots$}
\multiframe(42,11)(13.5,0){1}(10.5,10.5){\tiny{$\m_s$}}
\multiframe(42,0)(13.5,0){1}(10.5,10.5){\tiny{$\n_s$}}
\end{picture}
\label{Young}\ee and fulfills the Bianchi identity \be
\partial_{[\a} R_{\m_1 \n_1] \m_2 \n_2 \cdots \m_s \n_s} = 0\,.
\label{Bianchis}\ee

Conversely, given a tensor $R_{\m_1 \n_1 \m_2 \n_2 \cdots \m_s
\n_s}$ with the Young tableau symmetry (\ref{Young}) and fulfilling
the Bianchi identity (\ref{Bianchis}), there exists a ``potential"
$h_{\m_1 \m_2 \cdots \m_s}$ such that (\ref{DefR}) holds.  This
potential is determined up to a gauge transformation (\ref{gauges})
where the gauge parameter $\xi_{\m_2 \cdots \m_s}$ is unconstrained
\cite{Olver}.

\subsection{Equations of motion} The trace conditions on the gauge parameter for spins
$\geq 3$ are necessary in order to construct second-order invariants
-- and thus, in particular, gauge invariant second-order equations
of motion. One can show that the Fronsdal tensor $$ F_{\m_1 \m_2
\cdots \m_s} =\pa^\r \pa_\r h_{\m_1 \m_2 \cdots \m_s} -s
\pa_{(\m_1}\pa^\r h_{\m_2 \cdots \m_s)\r} + \frac{s(s-1)}{2}
\pa_{(\m_1\m_2}h_{\m_3 \cdots \m_s)\r}^{\hspace{1.2cm} \r} \,, $$
which contains only second derivatives of the potential, transforms
under a gauge transformation (\ref{gauges}) into the trace of the
gauge parameter
$$ F_{\m_1 \m_2 \cdots \m_s} \rightarrow F_{\m_1 \m_2 \cdots \m_s} +
\frac{(s-1)(s-2)}{2} \pa_{(\m_1 \m_2\m_3}\xi_{\m_4\cdots
\m_s)\r}^{\hspace{1.2cm} \r}\,,$$ and is thus gauge invariant when
the gauge parameter is requested to be traceless.  The Fronsdal
tensor is related to the curvature by the relation \be R_{ \m_1 \n_1
\m_2 \n_2 \cdots \m_s\n_s} \eta^{\n_1\n_2}=-\frac{1}{2}F_{\m_1\m_2
[\m_3 [\cdots [\m_s,\n_s] \cdots ]\n_3]}\,.\label{FR}\ee

In the absence of sources the equations of motion that follow from
the variational principle written in the main text (\ref{actions})
are \be G_{\m_1 \m_2 \cdots \m_s} = 0 \,,\ee where the ``Einstein"
tensor is defined as \be G_{\m_1 \m_2 \cdots \m_s} = F_{\m_1 \m_2
\cdots \m_s} -\frac{s(s-1)}{4} \eta_{(\m_1\m_2}F_{\m_3\cdots
\m_s)\r}^{\hspace{1.2cm} \r}\,.\label{eomsp}\ee They obviously imply
\be R_{ \m_1 \n_1 \m_2 \n_2 \cdots \m_s\n_s}
\eta^{\n_1\n_2}=0\label{eoms}\,,\ee and the inverse implication is
true as well \cite{Bekaert:2003az}. Indeed, Eq. (\ref{eoms}) implies
that the Fronsdal tensor has the form $ F_{\m_1 \m_2 \cdots \m_s}
=\pa_{(\m_1 \m_2\m_3}\Sigma_{\m_4\cdots \m_s)}$, which can be made
to vanish by a gauge transformation with an unconstrained gauge
parameter (see \cite{Francia:2002pt} for a discussion of the
subtleties associated with the double tracelessness of the spin s
field $h_{\m_1 \cdots \m_s}$). The interest of the equations
(\ref{eoms}) derived from the Einstein equations is that they
contain the same number of derivatives as the curvature. Thus, they
are useful to exhibit duality, which rotates the equations of motion
and the cyclic identities on the curvature.

\subsection{Dual curvature} The dual of the curvature tensor is
defined by
$$S_{\m_1 \n_1 \m_2 \n_2 \cdots \m_s\n_s} =-\frac{1}{2}
\e_{\m_1\n_1 \r \s}R^{\r\s}_{~~\m_2 \n_2 \cdots \m_s\n_s} \,,$$ and,
as a consequence of the equations of motion (\ref{eoms}), of the
symmetry of the curvature  and of the Bianchi identity
(\ref{Bianchis}), it has the same symmetry as the curvature and
fulfills the equations $ S_{ \m_1 \n_1 \m_2 \n_2 \cdots \m_s\n_s}
\eta^{\n_1\n_2}=0 $, $
\partial_{[\a} S_{\m_1 \n_1] \m_2 \n_2 \cdots \m_s\n_s}=0$.

\subsection{Conserved charges}

Non vanishing conserved charges can be associated with the gauge
transformations (\ref{gauges}) that tend to Killing tensors at
infinity (``improper gauge transformations"). They can be computed
from the constraints \cite{BR} or equivalently from the knowledge of
their associated conserved antisymmetric tensors $k^{[\a\b]}_\xi$
introduced in \cite{Barnich:2005bn}, which generalize the
electromagnetic $F_{\mu\nu}$, the divergence of which vanishes in
the absence of sources. The corresponding charge is given by $
Q_\xi=\frac{1}{2} \int_S \star k_\xi\,_{[\a\b]}\, dx^\alpha \wedge
dx^\beta$, where the integral is taken at constant time, over the
$2$-sphere at infinity. The tensors $k^{[\a\b]}_\xi $ read \beq
k^{[\a\b]}_\xi&=&\pa^\a h^{\b\m_1 \cdots \m_{s-1}}\xi_{\m_1 \cdots
\m_{s-1}}+
\frac{(s-1)}{2}\pa^{\b}h_\r^{~\r\m_1\cdots \m_{s-2}}\xi^\a_{~\m_1 \cdots \m_{s-2}} \nonumber \\
&&+(s-1)\pa_\r h^{\r\a\m_1 \cdots \m_{s-2}}\xi^\b_{~\m_1 \cdots
\m_{s-2}}
-\frac{(s-1)^2}{2} \pa^{(\a}h_\r^{~\m_1\cdots \m_{s-2})\r}\xi^\b_{~\m_1 \cdots \m_{s-2}}\nonumber \\
&&- (\a \leftrightarrow \b) +\cdots\,, \nonumber \eeq where the
dots stand for terms involving derivatives of the gauge
parameters.

Of particular interest are the charges corresponding to  gauge
transformations that are ``asymptotic translations'', i.e.
$\xi^{\m_1 \cdots \m_{s-1}}\rightarrow_{r\rightarrow \infty}
\epsilon^{\m_1 \cdots \m_{s-1}}$ for some traceless constant tensor
$\epsilon^{\m_1 \cdots \m_{s-1}}$ . For these transformations, the
charges become, using Stokes' theorem and the explicit expression
for $ k^{[\a\b]}_\xi$,
$$ Q_\epsilon =\epsilon_{\m_1 \cdots \m_{s-1}}\int_V
G^{0\m_1 \cdots \m_{s-1}}d^3x .$$ As these charges are conserved for
any traceless $\epsilon_{\m_1 \cdots \m_{s-1}}$, the quantities
$P^{\m_1 \cdots \m_{s-1}}$ defined as the traceless parts  of $
\int_V G^{0\m_1 \cdots \m_{s-1}}d^3x$ are conserved as well. In the
spin 2 case, $P^\m$ is the energy-momentum 4-vector.

\subsection{Electric sources}

In the presence of only electric sources, the  equations of motion
read: \be G_{\m_1 \m_2 \cdots \m_s} + \  T_{\m_1 \m_2
\cdots \m_s} =0\,,\label{Einsts}\ee or equivalently \be R_{ \m_1
\n_1 \m_2 \n_2 \cdots \m_s\n_s} \eta^{\n_1\n_2}=\frac{1}{2} \
\bar{T}_{\m_1\m_2 [\m_3 [\cdots [\m_s,\n_s] \cdots ]\n_3]}\ee where
$\bar{T}_{\m_1\m_2 \cdots \m_s}=T_{\m_1\m_2 \cdots \m_s}-
\frac{s}{4} \eta_{(\m_1\m_2}T^{\prime}_{\m_3\cdots \m_s)}\,$
generalizes the energy momentum tensor and primes denote traces,
$T^{\prime}_{\m_3\cdots \m_s}=T_{\m_1\cdots \m_s}\eta^{\m_1\m_2}\,.$
The curvature tensor has the Young symmetry (\ref{Young}) and
fulfills the Bianchi identity (\ref{Bianchis}), as in the case
without sources.

On the other hand, while the trace of the dual curvature tensor
still vanishes, the latter has no longer the Young symmetry
(\ref{Young}) and its Bianchi identity gets modified as well. The
new symmetry is described by the Young tableau
 \be
\begin{picture}(85,15)(0,2)
\multiframe(0,11)(13.5,0){1}(10.5,10.5){\tiny{$\m_1$}}
\multiframe(0,0)(13.5,0){1}(10.5,10.5){\tiny{$\n_1$}}
\multiframe(31,11)(13.5,0){1}(10.5,10.5){\tiny{$\m_2$}}
\multiframe(31,0)(13.5,0){1}(10.5,10.5){\tiny{$\n_2$}}
\multiframe(42,11)(13.5,0){1}(19.5,10.5){$\cdots$}
\multiframe(42,0)(13.5,0){1}(19.5,10.5){$\cdots$}
\multiframe(62,11)(13.5,0){1}(10.5,10.5){\tiny{$\m_s$}}
\multiframe(62,0)(13.5,0){1}(10.5,10.5){\tiny{$\n_s$}}
\put(15,7.5){$\otimes$}
\end{picture}\,, \label{Young1}
\ee as the dual curvature now satisfies $ S_{ [\m_1 \n_1 \m_2] \n_2
\cdots \m_s\n_s}=-\frac{1}{6}\ \e_{\m_1 \n_1\m_2
\r}\bar{T}^{\r}_{~\n_2  [\m_3 [\cdots [\m_s,\n_s] \cdots ]\n_3]}$,
while the Bianchi identity becomes $\partial_{[\a} S_{\m_1 \n_1]
\m_2 \n_2 \cdots \m_s\n_s} =\frac{1}{3}\
\e_{\a\m_1\n_1\r}\bar{T}^{\r}_{~[\m_2  [\m_3 [\cdots [\m_s,\n_s]
\cdots ]\n_3] \n_2]}$.

\section{Gauge invariances of higher spin action (with magnetic terms)}
\setcounter{equation}{0} \label{appendixGI}

We prove in this appendix that the action (\ref{actions}) is
invariant both under the standard symmetries of the spin s gauge
field and under displacements of the Dirac string.

To that end, we first observe that the first term in the action
(\ref{actions}) is invariant under the following shifts of $Y^{\m
\n}_{\; \; \; \; ~\a_1\cdots \a_{s-1}}$ : \be \d Y^{\m
\n}_{\; \; \; \;~ \a_1\cdots \a_{s-1}} =
\partial_\r\delta^\m_{(\a_1} z^{\n \r}_{ \; \; \; \; \a_2\cdots
\a_{s-1})}- \partial_\r\delta^\n_{(\a_1} z^{\m \r}_{ \; \; \; \;
\a_2\cdots \a_{s-1})}+
\partial_{(\a_1} z^{\m \n}_{ \; \; \; \; \a_2\cdots \a_{s-1})} \,,\label{gauge1s}\ee
where $z^{\m\n}_{ \; \; \; \; \a_1\cdots \a_{s-2}}  = z^{[\m\n]}_{
\; \; \; \; \a_1\cdots \a_{s-2}} =z^{\m\n}_{ \; \; \; \;
(\a_1\cdots \a_{s-2})} $  is an arbitrary traceless tensor that
satisfies $\eta^{\a_1[\l}z^{\m\n]}_{ \; \; \; \; \;\a_1\cdots
\a_{s-2}} =0$  when $s>2$. Under this transformation, $\bar{Y}^{\m
\n}_{\; \; \; \; ~\a_1\cdots \a_{s-1}}$ transforms as $ \d
\bar{Y}^{\m \n}_{~\; \; \; \; \a_1\cdots
\a_{s-1}}=\partial_{(\a_1} z^{\m \n}_{ \; \; \; \; \a_2\cdots
\a_{s-1})} $,  which makes it obvious that the curvature and the
Fronsdal tensor are invariant under (\ref{gauge1s}).

The transformation (\ref{gauge1s}) can be conveniently written \be
\d Y^{\m \n}_{~\; \; \; \; \a_1\cdots \a_{s-1}} =\e^{\m \n \r
\s} \partial_\r a_{\s \a_1 \a_2 \cdots \a_{s-1}}
\,,\label{gauge3s} \ee where $a_{\s \a_1 \a_2\cdots \a_{s-1}}
= - a_{ \a_1 \s  \a_2\cdots \a_{s-1}}=a_{\s \a_1 (
\a_2\cdots \a_{s-1})} $ is given by \be a_{\s \a_1 \a_2\cdots
\a_{s-1}} = \frac{1}{2} \e_{\s \b \g\a_1 } z^{\b \g}_{ ~~~
\a_2\cdots \a_{s-1} }\,,\ee is traceless and satisfies $a_{[\s
\a_1 \a_2] \a_3\cdots \a_{s-1}} =0$ when $s>2$.

Direct computation shows that the gauge transformation
(\ref{gauges}) of the spin s field acts on $Y^{\m \n}_{\; \; \;
\; ~\a_1\cdots \a_{s-1}} $ as the transformation (\ref{gauge3s})
with \beq a_{\r \s ( \a_1\cdots \a_{s-2})}
&=&-2\frac{(s-1)}{s}
\pa_{[\r} \xi_{\s ]\a_1\cdots \a_{s-2}} \nonumber \\
&&+\frac{(s-1)(s-2)}{s^2}\Big[\eta_{\r(\a_1}\pa^{\l}\xi_{\a_2\cdots
\a_{s-2})\l\s}-\eta_{\s (\a_1}\pa^{\l}\xi_{\a_2\cdots
\a_{s-2})\l\r}\Big]\nonumber \,.\eeq It follows from this fact and
the conservation of the energy-momentum tensor that the action
(\ref{actions}) is invariant under the standard gauge transformation
(\ref{gauges}) of the spin s field.

The displacements of the Dirac string change $\Phi^{\m \n}_{\; \; \;
\; \a_1\cdots \a_{s-1}} $ as $ \d \Phi^{\m \n}_{\; \;\; \;
\a_1\cdots \a_{s-1}} = k^{\m \n}_{\; \;\; \; \a_1\cdots \a_{s-1}} $
where $\pa_\m k^{\m \n}_{\; \;\; \; \a_1\cdots \a_{s-1}} =0\,.$ The
latter equation implies that $ k^{\m \n}_{\; \;\; \; \a_1\cdots
\a_{s-1}} =\pa_\l K^{\m \n\l}_{\; \;\; \; ~~\a_1\cdots \a_{s-1}} $,
where $K^{\m \n\l}_{\; \;\; \; ~\a_1\cdots \a_{s-1}}=K^{[\m
\n\l]}_{\; \;\; \; ~~\a_1\cdots \a_{s-1}}$. Let $\hat{K}^{\m
\n\l}_{\; \;\; \; ~\a_1\cdots \a_{s-1}}$ be the part of $K^{\m
\n\l}_{\; \;\; \; ~\a_1\cdots \a_{s-1}}$ that is traceless in
$\a_1\cdots \a_{s-1}$; it can be  decomposed as  $$ \hat{K}^{\m
\n\l}_{\; \;\; \; ~~\a_1\cdots \a_{s-1}} =x^{\m \n\l}_{\; \;\; \;
~~\a_1\cdots \a_{s-1}} +\d^{[\l}_{(\a_1} y^{\m\n]}_{~~\a_2\cdots
\a_{s-1})}\,,$$ where $x^{\m \n\l}_{\; \;\; \; ~~\a_1\cdots
\a_{s-1}}$ and $y^{\m\n}_{~~\a_2\cdots \a_{s-1}}$ satisfy \beq x^{\m
\n\l}_{\; \;\; \; ~~\a_1\cdots \a_{s-1}}=x^{[\m \n\l]}_{\; \;\; \;
~~\a_1\cdots \a_{s-1}}=x^{\m \n\l}_{\; \;\; \; ~~(\a_1\cdots
\a_{s-1})}\,,\
 x^{\m \n\l}_{\; \;\; \; ~~\a_1\cdots \a_{s-1}}\d_\l^{\a_1}=0 \,, \nonumber \\
y^{\m\n}_{~~\a_2\cdots \a_{s-1}}=y^{[\m\n]}_{~~\a_2\cdots
\a_{s-1}}= y^{\m\n}_{~~(\a_2\cdots \a_{s-1})}\,, \
y^{\m\n}_{~~\a_2\cdots
\a_{s-1}}\d_\n^{\a_1}=0\,, \nonumber \\
\eta^{\a_1[\l}y^{\m\n]}_{~~~\a_1 \cdots
\a_{s-1}}=-\frac{(s-1)}{2}\ x^{\m \n\l}_{\; \;\; \; ~~\a_1\cdots
\a_{s-1}}\eta^{\a_1\a_2}\,, y^{\m\n}_{~~~\a_1 \cdots
\a_{s-1}}\eta^{\a_1\a_2}=0\,. \nonumber \eeq For the action to be
invariant under displacements of the string, the variation of
$\Phi^{\m \n}_{\; \; \; \; \a_1\cdots \a_{s-1}} $ has to be
supplemented with an appropriate transformation of $h_{\a_1\cdots
\a_{s}} $. This transformation reads $ \d h_{\a_1\cdots
\a_{s}}=\frac{1}{6} \ \e_{\m\n\l(\a_1}\ x^{\m\n\l}_{~~~~\a_2\cdots
\a_{s})}\,.$ Indeed, when one performs both variations, $Y^{\m
\n}_{\; \; \; \; ~\a_1\cdots \a_{s-1}} $ transforms as in
(\ref{gauge1s}), so the first term in the action is invariant.
Furthermore, the electric coupling term is invariant as well
because the support of the variation of the spin s field does not
contain the electric worldlines.

The identities that follow from the invariance (\ref{gauge1s}) --
or (\ref{gauge3s}) -- of the first term  ${\cal L}$  in the action
may be written conveniently in terms of \be A^{\s \g_1\cdots
\g_{s-1}}=\e^{\s\m\n\l }
\partial_\l \left(\frac{\partial {\cal L}}{\partial Y^{\m \n}_{\;
\; \; \; ~\g_1\cdots \g_{s-1}}}\right)\,, \ee and its trace $A^{
\prime\g_2\cdots \g_{s-1}}=A^{\s \g_1\cdots
\g_{s-1}}\eta_{\s\g_1}$ . They read \be 0=A^{\s \g_1\cdots
\g_{s-1}}-A^{(\g_1 \g_2\cdots
\g_{s-1})\s}-\frac{s-2}{s}\Big(\eta^{\s(\g_1}A^{ \prime\g_2\cdots
\g_{s-1})}-\eta^{(\g_1\g_2}A^{ \prime\g_3\cdots \g_{s-1})\s}\Big)\,.
\ee Using these identities, one checks the following useful
relations: \beq G^{\g_1 \cdots \g_{s}}=\frac{\delta {\cal L}}{\delta
h_{ \g_1 \cdots \g_{s}}} &=&A^{(\g_1\g_2 \cdots
\g_{s})}+\frac{(s-1)(s-2)}{2s}
\eta^{(\g_1\g_2}A^{\prime \g_3\cdots \g_{s})}\\
&=&A^{\g_1\g_2 \cdots
\g_{s}}+\frac{(s-1)(s-2)}{2s}\eta^{(\g_2\g_3}A^{\prime \g_4\cdots
\g_{s})\g_1}\,, \label{useful} \eeq which will be used in the
following appendix.

\section{Quantization condition for electric and magnetic higher spin sources}
\setcounter{equation}{0} \label{QCHS}

We work out explicitly in this appendix the quantization condition
for higher spins.  As for lower spins, this is achieved by
integrating the constraints expressing the unobservability of the
Dirac string.

In the gauge $y^0 = \l$, these constraints read $$ \pi_m = -2 N y'^n
\, f_{\g_1 \cdots \g_{s-1}}(v)\frac{\partial {\cal L}}{\partial
Y^{mn}_{\; \; \;\; \; ~\g_1 \cdots \g_{s-1}}} \,.$$ In the
quantum theory, the wave functional $\psi$ must thus fulfill $$
\frac{\hbar}{i} \frac{\delta \psi}{\delta y^m(\s)} = -2N y'^n \,
f_{\g_1 \cdots \g_{s-1}}(v)\frac{\partial {\cal L}}{\partial
Y^{mn}_{\; \; \; \; \; ~\g_1 \cdots \g_{s-1}}} \psi \,.$$
Integrating this equation along a path that encloses an electric
source, one finds the following variation of the phase of the wave
functional  $$ \Delta \Psi = - \frac{ N }{\hbar} \, f_{\g_1 \cdots
\g_{s-1}}(v)\int \frac{\partial {\cal L}}{\partial Y^{mn}_{\;
\; \; \; \; ~\g_1 \cdots \g_{s-1}}} \left(\dot{y}^m y'^n - \dot{y}^n
y'^m \right) d \s d \l \,,$$ where the integral is taken on the
two-dimensional surface enclosing the electric source. Using the
Gauss theorem, this can be converted into a volume integral, $$
\Delta \Psi =  -\frac{N  }{\hbar} \, f_{\g_1 \cdots \g_{s-1}}(v)\int
d^3 x\ \e^{mnp} \partial_p \left(\frac{\partial {\cal L}}{\partial
Y^{mn}_{\; \; \; \;\;~ \g_1 \cdots \g_{s-1}}} \right)\,. $$
Using the relations (\ref{useful}), one checks that $$ \e^{mnp}
\partial_p \left(\frac{\partial {\cal L}}{\partial Y^{mn}_{\;~\; \;
\; \; \g_1 \cdots \g_{s-1}}} \right) = \frac{\delta {\cal L}}{\delta
h_{0 \g_1 \cdots \g_{s-1}}} +\cdots\,,$$ where the dots stand for
terms of the form $\eta^{(\g_1\g_2}X^{\g_3 \cdots \g_{s-1})}$. Upon
use of the Einstein equations $G^{0 \g_1 \cdots \g_{s-1}} =
-T^{0 \g_1 \cdots \g_{s-1}}$, the variation of the phase
becomes,  $ \Delta \Psi = \frac{N }{\hbar} \, f_{\g_1 \cdots
\g_{s-1}}(v) \int d^3 x\ T^{0 \g_1 \cdots \g_{s-1}} = \frac{ MN
}{\hbar} \, f_{\g_1 \cdots \g_{s-1}}(v)f^{\g_1 \cdots \g_{s-1}}(u)
$.  For the wave functional to be single-valued, this should be a
multiple of $2 \pi$.  This yields the quantization condition \be
\frac{MN  }{2 \pi \hbar} \, f_{\g_1 \cdots \g_{s-1}}(v)f^{\g_1
\cdots \g_{s-1}}(u)= n , \; \; \; n \in \Bbb{Z} \,.\ee Introducing
the conserved charges $P^{\g_1 \cdots \g_{s-1}}$, $Q^{\g_1 \cdots
\g_{s-1}}$, this can be rewritten as \be \frac{1  }{2 \pi \hbar} \,
Q_{\g_1 \cdots \g_{s-1}}(v)P^{\g_1 \cdots \g_{s-1}}(u) \in \Bbb{Z}
\,.\ee

\section{Rotation Symmetry and Charge Quantization for an Electromagnetic
Magnetic Pole} \setcounter{equation}{0} \label{appendixB}

\subsection{The Dirac string and rotations} \label{rots}

A monopole is spherically symmetric, yet the vector potential is
not manifestly so, and when rotated, the string will move to a new
location. As is well known, spherical symmetry is recovered by
supplementing the naive rotation generated by the usual Killing
vectors by a gauge transformation that moves the string back to
its original location. Let us see how this works explicitly.

The usual Killing vectors which generate SO(3) rotations on the
two-sphere are given by \bea \label{killing1} \xi_X & = &
-\sin\phi \, \frac{\6}{\6 \theta} - \cos\phi \cot\theta \,
\frac{\6}{\6 \phi} \,, \nn \xi_Y & = & \cos\phi \, \frac{\6}{\6
\theta} - \sin\phi \cot\theta \, \frac{\6}{\6 \phi} \,, \nn \xi_Z
& = & \frac{\6}{\6 \phi} \,. \eea They satisfy the algebra \be
\left[ \xi_A, \xi_B \right] = -\epsilon_{ABC} \xi_C\,. \ee Under
an infinitesimal rotation generated by one of these vectors, the
vector potential changes by the Lie derivative along the vector
field: \be \mathcal{L}_{\xi_B} A_i = \xi^j_B \6_j A_i + A_j \6_i
\xi^j_B = \xi^j_B F_{ji} +
\partial_i(\xi^j_B A_j)\,. \ee For the monopole vector potential
(\ref{vecpot}) this Lie derivative is non-zero and hence usual
rotational symmetry would appear to be broken. However, we can undo
its effect if we accompany the Lie derivative by a gauge
transformation $A_i \rightarrow A_i + \partial_i \Lambda$ such that
the total change is \be \label{vara} \delta_{\xi_B} A_i =
\mathcal{L}_{\xi_B} A_i +
\partial_i \Lambda_B =0 \,.\ee The gauge transformations that we find
associated with the three Killing vector fields are \bea
\label{lambdas} \Lambda_X & = & - \frac{g}{4\pi} \cos\phi
\sin\theta - \xi^j_X A_j = \frac{gk}{4\pi} \cot\theta \, \cos\phi
- \frac{g}{4\pi} \frac{\cos\phi}{\sin\theta} \nn \Lambda_Y & = & -
\frac{g}{4\pi} \sin\phi \sin\theta - \xi^j_Y A_j = \frac{gk}{4\pi}
\cot\theta \, \sin\phi - \frac{g}{4\pi}
\frac{\sin\phi}{\sin\theta} \nn \Lambda_Z & = & - \frac{g}{4\pi}
\cos\theta - \xi^j_Z A_j = - \frac{gk}{4\pi} \,.\eea We note that
$\Lambda_Z$ is a constant gauge parameter, yet it needs to be
considered in order that the new rotation generators to be defined
below satisfy the algebra.

We can combine the gauge transformations $\Lambda_B$ and the naive
rotation operators $\xi_B $ to create new rotation operators
$\hat\xi_B=\xi_B+\Lambda_B \frac{\partial}{\partial \lambda}$ by
using the fact that the generator of gauge transformations is the
electric charge $q$, which is in turn conjugate to the coordinate on
the U(1) fiber which we call $\lambda$, so we can write: \bea
\label{gg2} && \hat\xi_X = -\sin\phi \, \frac{\6}{\6 \theta} -
\cos\phi \cot\theta \, \frac{\6}{\6 \phi} + \left( \frac{g k}{4\pi}
\cos\phi \cot\theta - \frac{g}{4\pi}
\frac{\cos\phi}{\sin\theta}\right) \frac{\partial}{\partial \lambda}
\,, \nn & &\hat\xi_Y = \cos\phi \, \frac{\6}{\6 \theta} - \sin\phi
\cot\theta \, \frac{\6}{\6 \phi} + \left( \frac{g k}{4\pi} \sin\phi
\cot\theta - \frac{g}{4\pi} \frac{\sin\phi}{\sin\theta}
\right)\frac{\partial}{\partial \lambda} \,, \nn & &\hat\xi_Z =
\frac{\6}{\6 \phi} - \frac{g k}{4\pi}\frac{\partial}{\partial
\lambda} \,. \eea These operators satisfy the SO(3) algebra just as
(\ref{killing1}).

\subsection{The field angular momentum}

Let us now consider an electric charge $q$ of mass $m$ placed at a
point $\vec{a}$ in the background field of a magnetic pole $g$
placed at the origin. The background magnetic and dynamical electric
fields are \be \vec{B} = \frac{g \, \vec{x}}{ 4 \pi |x|^3}\,, \qquad
\vec{E} = \frac{q \, (\vec{x}-\vec{a}) }{ 4 \pi |x-a|^3}\,. \ee As
is well known, the angular momentum stored in the electromagnetic
field does not vanish. It can be calculated from the standard
symmetric energy-momentum tensor of the electromagnetic field and is
given by \be \label{belin} {\vec L}_{field} = \int_V \vec{x} \times
(\vec{E} \times \vec{B}) = \frac{q g}{4\pi} \frac{\vec{a}}{ |a|}
\,.\ee This expression is independent of the magnitude of $\vec{a}$,
though it depends on its direction, and therefore the limit
$\vec{a}\rightarrow 0$ cannot be taken continuously. For example, if
we place the charge above the pole the angular momentum will be
negative, whereas it will be positive if placed below. Whatever the
sign, if we assume the angular momentum to be quantized in integer
multiples of $\hbar/2$, we obtain the quantization condition \be
\frac{qg}{2\pi} \in \Bbb{Z} \, .\ee This was first noted by Fierz
and is the statement that quantization of the field angular momentum
results in the Dirac quantization condition.  The fact that one may
obtain half integer angular momentum without fermions in sight has
been lucidly analyzed in \cite{Goldhaber:1976dp}.

\subsection{Angular momentum of a test electric charge in the
background of a magnetic monopole}

One can get direct access to the angular momentum in the field
differently, by analyzing the angular momentum of a test electric
charge in the background of a magnetic pole.

The magnetic field of a magnetic monopole of strength $g$, \be B^r
= \frac{g}{4\pi r^2}\,, \ee can be associated with the one form
vector potential \be \label{vecpot}A = \frac{g}{4\pi} (k -
\cos\theta)\, d\phi\,, \ee which satisfies $B=\nabla\times A$, for
$k$ an arbitrary constant, away from the Dirac string. When $k=1$
the Dirac string lies along the negative $z$-axis while when
$k=-1$ the Dirac string lies along the positive $z$-axis. The
value of $k$ can be modified to $\tilde k$ by a gauge
transformation with parameter $g(\tilde k - k)\phi/4\pi$. {}For
values of $k \not= \pm 1$, there are two strings, one along the
positive $z$-axis and one along the negative $z$-axis, with
strengths that add up to one.

By demanding that the velocity should transform as a vector under
rotations, the angular momentum operator  \be \label{op} \vec{J} = m
\vec{x} \times \dot{\vec{x}} + \frac{qg}{4\pi}
\frac{\vec{x}}{|x|}\,, \ee was derived in \cite{Coleman:1982cx}.

There is an alternative way to arrive at this result which we
believe is worth recalling. Using the standard Noether procedure for
the action \be \label{lag} S = \int \left( \frac{1}{2} m \dot x_i
\dot x^i - q A_i \dot x^i \right) dt \ee we obtain that for a
variation $\delta x$ the action changes by \be \label{var1}\delta S
= \int \frac{d}{dt} \left[ (m \dot x_i - q A_i) \delta x^i
\right]\,. \ee on-shell, i.e. when the equations of motion hold. The
transformation will generate a conserved charge if the action is
invariant off-shell up to the integral of a total derivative. Let us
consider what the case is for a rotation generated by $\delta x^i =
\xi_B^i$. The first term in (\ref{lag}) is the norm of a vector and
therefore manifestly invariant under rotations. The variation of the
second term becomes \bea \delta S & = & -q \int \left( \dot x^i
\delta_{\xi_B} A_i + A_i \delta_{\xi_B} \dot x^i\right) \nn & = & -q
\int \left( {\mathcal L}_{\xi_B} A_i \right) \dot x^i \,,\eea where
we have used $\delta_{\xi_B}A_i = (\partial_j A_i) \,
\delta_{\xi_B}x^j$. If, as is the case here (see (\ref{vara})), the
Lie derivative of $A_i$ is given by a total derivative ${\mathcal
L}_{\xi_B} A_i = -\partial_i \Lambda_B$ this becomes \be
\label{var2} \delta S = q \int \frac{d}{dt} \left[ \Lambda_B \right]
\,.\ee We can now equate equations (\ref{var1}) and (\ref{var2}) to
obtain the charge $J_B$ conserved under rotations \be J_B = \left( m
\dot x_i -q A_i \right) \xi_B^i - q \Lambda_B\,. \ee In fact the
$\Lambda_B$ are the gauge transformations we computed in
$(\ref{lambdas})$. For example, for rotations about the z-axis, we
obtain that \be J_Z = m \dot \phi + \frac{qg}{4 \pi} \cos\theta
\,,\ee and in general the result (\ref{op}).

{}From the expression (\ref{op}), it is clear that the angular
momentum does not vanish when the particle is at rest. This apparent
paradox is explained by comparing (\ref{op}) with (\ref{belin}). The
extra piece is the angular momentum of the electromagnetic field,
which must be included because the electric field of the test
particle is dynamical and hence, the angular momentum of the
electromagnetic field changes with time. Only the sum of the
standard orbital angular momentum and the field angular momentum is
conserved.

Expressing the angular momentum in terms of the (non gauge
invariant) conjugate momenta $p_\phi=m \dot\phi - qA_\phi$ one
obtains, for the angular momentum about the z-axis, \be J_Z = p_\phi
+ k \frac{qg}{4\pi}\label{improved}\,.\ee In the gauge $k=1$ and
when the particle is on the positive $z$-axis, the field angular
momentum is just equal to the extra piece appearing in $J_Z$,
besides the usual $p_\phi$. This is a way to identify the field
angular momentum if one knows the total angular momentum, as it is
the case for gravity. Note that in the gauge $k=0$, $p_\phi$ is
equal to the total angular momentum.

It has been shown in \cite{Goldhaber:1976dp,Coleman:1982cx}, that
the difference $J_Z-qg/4\pi$ has integer eigenvalues.  We may
rewrite this assertion in the notation used here as the statement
that in the gauge $k=1$ the operator $p_\phi=\frac{\hbar}{i}
\frac{\partial}{\partial\phi}$ has integer eigenvalues and therefore
the wave function is periodic in $\phi$.

However the periodicity of the wave function depends on $k$.  This
can be traced to the fact that differentiating with respect to
$\phi$ \be p_\phi = \frac{\hbar}{i} \frac{\partial}{\partial\phi}
\vert_{\lambda} \ee at constant $\lambda$ (where $\lambda$ is the
coordinate along the U(1) fiber) is not invariant under the gauge
transformation \be \label{gautrans}\lambda \rightarrow \lambda -
\frac{g \phi}{4\pi} \,,\ee which brings $k$ from 1 to 0. As
(\ref{gautrans}) itself shows the wave function picks up a phase
factor $e^{-i\frac{q g}{4 \pi}\phi}$ which shows in turn, that in
the gauge $k=0$ the wave function is antiperiodic in $\phi$ when
$qg/4\pi$ is half integer. This anti-periodicity is permissible
because in the gauge $k=0$ an infinite line is removed from $R^3$
and the resulting configuration space for the electron is therefore
not simply connected.  The infinite line is formed by two strings of
half strength that come into the magnetic pole along the positive
and negative sides of the $z$-axis.

There is another interesting manifestation of the lack of simple
connectedness due to the infinite line formed by the two strings
when $k = 0$.  It is the following: the closed path traveled in
configuration space when the string sweeps out a closed surface
around the electric charge is not contractible to the identity,
whereas it is so when the string is half infinite (see
\cite{Deser:1997mz}). For this reason the string wave function
changes sign after the turn and one gets the same quantization
condition as when $k=1$ in which case the path is contractible and
the wave function returns to its original value.

\end{appendix}

\end{document}